# Disordered skyrmion phase stabilized by magnetic frustration in a chiral magnet

K. Karube,[1]*† J. S. White,[2]† D. Morikawa,[1] C. D. Dewhurst,[3] R. Cubitt,[3] A. Kikkawa,[1] X. Z. Yu,[1] Y. Tokunaga,[4] T. Arima,[1,4] H. M. Rønnow,[5] Y. Tokura,[1,6] Y. Taguchi[1]

[1]RIKEN Center for Emergent Matter Science (CEMS), Wako 351-0198, Japan.

[2]Laboratory for Neutron Scattering and Imaging (LNS), Paul Scherrer Institute (PSI), CH-5232 Villigen, Switzerland.

[3]Institut Laue-Langevin (ILL), 71 avenue des Martyrs, CS 20156, 38042 Grenoble cedex 9, France.

[4]Department of Advanced Materials Science, University of Tokyo, Kashiwa 277-8561, Japan.

[5]Laboratory for Quantum Magnetism (LQM), Institute of Physics, École Polytechnique Fédérale de Lausanne (EPFL), CH-1015 Lausanne, Switzerland.

[6]Department of Applied Physics, University of Tokyo, Bunkyo-ku 113-8656, Japan.

*Corresponding author. Email: kosuke.karube@riken.jp

†These authors contributed equally to this work

**Abstract**

Magnetic skyrmions are vortex-like topological spin textures often observed to form a triangular-lattice skyrmion crystal in structurally chiral magnets with Dzyaloshinskii-Moriya interaction. Recently $\beta$-Mn structure-type Co-Zn-Mn alloys were identified as a new class of chiral magnet to host such skyrmion crystal phases, while $\beta$-Mn itself is known as hosting an elemental geometrically frustrated spin liquid. Here we report the intermediate composition system $Co_7Zn_7Mn_6$ to be a unique host of two disconnected, thermal-equilibrium topological skyrmion phases; one is a conventional skyrmion crystal phase stabilized by thermal fluctuations and restricted to exist just below the magnetic transition temperature $T_c$, and the other is a novel three-dimensionally disordered skyrmion phase that is stable well below $T_c$. The stability of this new disordered skyrmion phase is due to a cooperative interplay between the chiral magnetism with Dzyaloshinskii-Moriya interaction and the frustrated magnetism inherent to $\beta$-Mn.

**Introduction**

Magnetic spin systems in solids exhibit a rich variety of ordering patterns, dependent on the microscopic interactions, anisotropy, lattice form, etc. Among the various orders, non-collinear or non-coplanar ones with vector or scalar spin chirality attract considerable attention due to the associated collective properties they may generate, such as multiferroic or emergent electromagnetic responses. Magnetic skyrmions (*1-4*), vortex-like spin swirling textures characterized by an integer topological charge, are a quintessential example of non-coplanar magnetic structures. Thus far, they have been observed or predicted in various systems (*3–10*), with their origins attributed to several microscopic mechanisms, such as competition between Dzyaloshinskii-Moriya interaction (DMI) and ferromagnetic exchange interaction, magnetic frustration (*11-13*), Fermi surface effect (*14*), and magnetic dipolar interaction (*15*). In particular, a finite DMI can arise due to broken inversion symmetry either at interfaces of thin-film layers (*5, 6*) or in bulk materials with chiral or polar structures (*3, 4, 7-10*). In the chiral magnets, the effect of the DMI is to gradually twist otherwise ferromagnetically coupled moments to form a helical ground state with a mesoscale periodicity described by a single magnetic propagation vector (*q*). An applied magnetic field may induce a triangular-lattice skyrmion crystal (SkX), which is often described as a triple-*q* structure with the *q*-vectors displaying mutual angles of 120°. Such SkX states are stabilized by thermal fluctuations and confined to a narrow region near the helimagnetic transition temperature $T_c$ (*2, 3*).

Recently, DMI-based skyrmions have been observed in Co-Zn-Mn alloys with the *β*-Mn-type chiral cubic structure (*9*), where the unit cell contains 20 atoms distributed over two inequivalent crystallographic sites (8*c* and 12*d* Wyckoff sites, the inset of Fig. 1A). $Co_{10}Zn_{10}$, one end-member of a solid solution $(Co_{0.5}Zn_{0.5})_{20-x}Mn_x$ ($0 \leq x \leq 20$) with the $\beta$-Mn structure (*16*), shows a helical ordering of Co spins (periodicity $\lambda \sim 185$ nm) below $T_c \sim 460$ K, with $T_c$ decreasing as the partial substitution of Mn proceeds. In $Co_8Zn_8Mn_4$ ($T_c \sim 300$ K), a SkX state created close to $T_c$ can persist as a metastable state over a very wide temperature and magnetic field region upon a field cooling (FC), accompanied by a lattice-form transformation of the SkX at low temperatures (*17, 18*). The other end-member $Mn_{20}$ ($\beta$-Mn itself) is well-known to display no transition to magnetic long range order (*19, 20*) due to the strong geometrical frustration of antiferromagnetic interactions in the three-dimensional hyper-kagome network of the 12*d* sites (*21*). Lightly doped $\beta$-Mn alloys have been found to exhibit spin glass states (*19, 21*) or complex incommensurate antiferromagnetic states with non-coplanar structure (*22*).

Fig. 1A shows the *T* (temperature) - *x* (Mn composition) phase diagram connecting $Co_{10}Zn_{10}$ and $\beta$-Mn according to $(Co_{0.5}Zn_{0.5})_{20-x}Mn_x$ as revealed in the present study (see Fig. S2). A spin glass phase symptomatic of frustrated magnetism is found to exist at low temperatures, and over a wide range of *x*. For $3 \leq x \leq 7$ the spin glass phase invades the helical phase and displays a typical reentrant spin glass behavior (*23, 24*), indicating a microscopic coexistence of the two states. To investigate the influence of frustration on the helical and topological spin textures, we focused on $Co_7Zn_7Mn_6$ ($x = 6$, indicated with a pink arrow in Fig. 1A) with $T_c \sim 160$ K and spin glass transition temperature $T_g \sim 30$ K, and performed measurements of small angle neutron scattering (SANS), magnetization, ac susceptibility, and Lorentz transmission electron microscopy (see Materials and Methods).

As summarized in the *T-H* (magnetic field) phase diagram in Fig. 2A, two distinct, equilibrium skyrmion phases are found; one is a conventional SkX phase slightly below $T_c$, and the other is a novel disordered skyrmion (DSk) phase near $T_g$. This new phase is quenched as a metastable state down to zero field in the field-decreasing process (Fig. 2B). To substantiate our findings in what follows, we summarize the relation between the real-space magnetic structures and the corresponding SANS patterns in Fig. 1B.

**Results**

**SANS measurements of conventional skyrmion phase at high temperatures**

At high temperatures just below $T_c$, a conventional SkX state is observed (Fig. 1C). The SANS pattern changes from 4 spots to 12 spots upon the application of magnetic field in the $H \parallel$ beam geometry at 146 K, indicating the transition from a helical multi-domain state to a triangular-lattice SkX state (see Fig. S4 for details). The SkX state coexists with a conical state (2 horizontal spots in the $H \perp$ beam geometry). As the temperature is lowered, the volume fraction of the SkX state relative to the conical state becomes smaller (see Figs. S4 and S5 for details). Consequently, the *T-H* phase diagram above 100 K is characterized by a typical SkX pocket near $T_c$ (green region in Fig. 2A) in an otherwise conical phase background.

**SANS and ac susceptibility measurements in a zero-field cooling**

Upon zero-field cooling (ZFC) the periodicity $\lambda$ of the helical state shrinks (from 110 nm to 73 nm), and the diffraction spots in the SANS pattern broaden significantly below ~

90 K providing evidence for growing magnetic disorder upon cooling (see Figs. S3D,E). In this ZFC process, ac susceptibility smoothly decreases below ~ 90 K and drops at ~ 30 K, the latter of which corresponds to a spin glass transition as evidenced by a strong frequency dependence of the transition temperature (see Figs. S3B,C). The disordering of the helical state toward low temperature and the subsequent spin glass transition are likely due to the enhancement of frustrated antiferromagnetic correlations of Mn spins which couple ferromagnetically with Co spins. The spin glass phase exists over a wide $x$ range with a nearly constant $T_g$ ~ 60 K in the $T$-$x$ phase diagram (Fig. 1A), suggesting a continuous existence of frustrated antiferromagnetic Mn spin correlations from $\beta$-Mn to $Co_7Zn_7Mn_6$.

**SANS measurements at low temperatures**

Figure 3 shows SANS results from magnetic field up and down scans at 50 K (see also Fig. S6). In the field-increasing run after ZFC (Fig. 3B), the pattern with 4 broad spots changes to a ring pattern above 0.1 T and finally SANS intensity disappears in the induced ferrimagnetic region (*25*). In the subsequent field-decreasing run, a ring pattern appears again and persists to zero field. The azimuthal angle dependent scattering intensity is plotted against magnetic field in Fig. 3C. Above 0.1 T in the field increasing run and over the whole field region in the field-decreasing run, intensities for directions close to <100> and <110> overlap completely, as expected for a homogeneous ring of scattering. Since ring-like patterns are also observed in the $H \perp$ beam geometry at 0.1 T (Fig. 3D) and even after the removal of magnetic field (Fig. 3E), we conclude the scattering intensity to be three-dimensionally distributed over a spherical shell in reciprocal space. This spherical

SANS pattern may be explained in terms of two possible scenarios; (i) a three-dimensionally disordered helical state (a topologically trivial state) or (ii) a three-dimensionally disordered skyrmion state (a topologically defined state) as illustrated in Fig. 1Bv. In the following sections, we show that the scenario (ii) is more plausible.

**SANS measurements in several warming processes**

To distinguish the above two possibilities [(i) and (ii)], we present SANS results obtained during the following two warming processes done subsequently to field sweepings at low temperatures.

First, in the zero-field warming (ZFW) process done after the field-decreasing run at 50 K (see Fig. S9), the ring pattern in the $H \parallel$ beam geometry changes to a pattern with 4 sharp spots (helical multi-domain state) above ~ 90 K. In a subsequent ZFC process returning from 130 K, the observed pattern at 50 K is no longer ring-like, but forms 4 broad spots, indicating the strongly irreversible nature of the low-temperature state.

Second, an important clue to the nature of the low-temperature state is provided by SANS data in the $H \perp$ beam geometry in the field-warming (FW) process at 0.025 T done subsequently to the field-decreasing run at 60 K (Fig. 4). As shown in Fig. 4B, in this FW process, the ring-like patterns observed in the $H \perp$ beam geometry also change to a 4 spot pattern above 90 K due to reduced disorder effect at elevated temperatures. Remarkably, while the intensity nearly parallel to the field ($q \parallel H$) is slightly stronger than that nearly perpendicular to the field ($q \perp H$) in the ring-like pattern at 60 K, the 2 vertical spots ($q \perp H$) become stronger than the 2 horizontal spots ($q \parallel H$) in the 4 spot pattern at 90 K and 110 K. If the 2 vertical spots were attributed to an ordered helical domain restored from its low temperature disordered state, the 2 horizontal spots (from the conical state) should remain stronger than the 2 vertical spots in this field-forced condition, which is in contradiction to the actual observation. On the other hand, this characteristic change in intensity distribution is reasonably explained if the 2 vertical spots at 90 K and 110 K arise from two-dimensional skyrmions, which are created with random orientation at low temperatures, persist as a superheated metastable state, and become more ordered with respect to the field direction above the order-disorder crossover region (grey hatching area in Fig. 2). The temperature-dependent intensities for $q \perp H$ and $q \parallel H$, and their ratio are presented in Fig. 4C and Fig. 4D, respectively. With increasing temperature up to 110 K, the intensity observed from the $q \parallel H$ region decreases and transfers to the $q \perp H$ region. This indicates that the broad intensity distribution around $q \parallel H$ at 60 K can be attributed not only to conical state, but also to vertically-oriented skyrmions, thus confirming that the skyrmions at low temperatures are of a *three*-dimensionally disordered form as illustrated in Fig. 1Bv.

Upon further increase of the temperature above 110 K, the intensity of the 2 vertical spots becomes weaker, which is interpreted as an eventual changeover of the metastable SkX state into the equilibrium conical state. Therefore, the aforementioned irreversibility is better interpreted in terms of a destruction of a metastable skyrmion state rather than, for example, the loss of a disordered helical state with strongly glassy character.

**Ac susceptibility measurements in several warming processes**

The above interpretation is further reinforced by ac susceptibility data shown in Fig. 5. Figure 5B shows the field dependence of the real-part of the ac susceptibility [$\chi'(H)$] at 110 K measured after FC and FW processes at 0.025 T as illustrated in Fig. 5A. After the FC1 process (blue circles) passing through the conventional SkX phase, $\chi'(H)$ shows an asymmetric shape with respect to the sign of $H$, which is a hallmark signature for a supercooled metastable SkX state surviving predominantly at positive-field region (*17, 28*). In the field-returning process from field-polarized ferrimagnetic regions, the conical state exhibits a symmetric shape of $\chi'(H)$ with larger values (black line) than the metastable SkX state. After the FW1 process (red line) passing through the low-temperature DSk phase, $\chi'(H)$ again shows an asymmetric shape similarly to the case of FC1, indicating that superheated metastable SkX state is realized above the crossover region.

In contrast to these behavior, $\chi'(H)$ is observed to be symmetric (Fig. 5D) as expected for a helical domain state, after both the FC2 process (yellow squares) that bypasses the conventional SkX phase, and the FW2 process (green line) that passes through the highly glassy region at 2 K but bypasses the DSk phase, as illustrated in Fig. 5C. Especially, the latter data indicate that the strongly glassy character at low temperatures is completely suppressed already at 110 K. Therefore, the asymmetric shape of $\chi'(H)$ in the FW1 process cannot be ascribed to a helical-domain state with the glassy character. These results clearly indicate that a novel equilibrium skyrmion phase, namely DSk phase, exists at low temperatures in addition to the conventional SkX phase just below $T_c$, and that the metastable SkX state is commonly realized by both superheating and supercooling, respectively.

The asymmetry of $\chi'(H)$ in the FW process passing through the DSk phase is diminished as the temperature is increased above the crossover region in a warming-field dependent manner: It is lost at lower temperature as the warming-field is decreased, as presented in Fig.S10 in detail. This behavior is also consistent with the interpretation that the asymmetry represents existence of the metastable skyrmions since they are generally known to become more fragile as the field is reduced (*17, 26-28*).

Taking the above SANS and ac susceptibility data in the warming processes into account, it is concluded that the spherical SANS pattern observed at low temperatures originates from (ii) disordered skyrmions, rather than (i) disordered helical state with strongly glassy character.

**SANS measurements below the spin glass transition temperature**

The DSk phase is found to exist at least below 60 K and above $T_g \sim 30$ K as an equilibrium state, on the basis of field-swept SANS measurements at different temperatures (see Fig. S8), as summarized in the *T-H* phase diagram in Fig. 2. The DSk phase is not stabilized below $T_g$ after ZFC as evidenced by the field-dependent SANS measurement at 20 K (Figs. S7A-C). On the other hand, the DSk phase is accessible by a FW process from temperatures below $T_g$, as demonstrated in Figs. S7D-F. Once created, the disordered skyrmions also persist below $T_g$ as a supercooled metastable state in the subsequent FC process.

**LTEM measurements**

Figure 6 shows real-space observations by Lorentz transmission electron microscopy (LTEM) on a thin-plate sample (thickness ~ 150 nm). In the field-increasing run at 135 K after ZFC, the transition from a helical state to a conventional SkX state is clearly observed (Fig. 6B). At the lower temperature of 50 K, a disordered helical state is observed as a distribution of elongated objects after ZFC (Fig. 6C). In contrast, under a magnetic field of 0.2 T, a number of closed, dot-like objects, assigned to skyrmions, are observed clearly. It is noted that the magnetic contrast of these skyrmions is weaker than at 135 K (Fig.6B) despite the fact that the local magnetization becomes larger, implying either a magnetic modulation along, or a slight inclination of the skyrmions from, the normal direction of the sample plate, consistent with the spherical distribution of $q$-vectors in SANS measurements. While the distribution of skyrmions displays positional disorder, they are identified to be topologically nontrivial as clearly indicated from the in-plane magnetization map derived from transport-of-intensity equation analysis of the LTEM images (Fig. 6D; see Materials and Methods for detailed procedures of the analysis). At 0.4 T in the ferrimagnetic state, the magnetic contrast disappears, and at 0 T after the subsequent removal of the field, the magnetic contrast due to a mixture of disordered skyrmions and disordered helices is observed. In the subsequent ZFW process, the disordered skyrmions coalesce to form disordered helices [with a number of topologically nontrivial defects (*29, 30*)] at 80 K and finally become an ordered helical state at 100 K. These temperature and field variations are fully consistent with the SANS results on the bulk crystal sample. In some materials, it is reported that the equilibrium skyrmion phase tends to expand toward lower temperatures in a thin-plate specimen with a thickness comparable to a skyrmion diameter (*4, 7, 8*). In the present Co-Zn-Mn alloys, however, the effect of a reduced sample thickness on the stability of the equilibrium SkX phase is relatively small: the SkX is observed only over a temperature range that amounts to approximately 10% of $T_c$ even for thin-plate samples (*9*). Thus, the present LTEM observation of disordered skyrmions at 50 K can hardly be explained as an effect of reduced dimensionality alone, but should be regarded as an inherent feature of the material.

**Discussion**

Here, we discuss the possible origin of the novel DSk phase. Skyrmion phases are generally believed to be stabilized with the help of either thermal (*3*) or quantum critical (*31, 32*) fluctuations. As discussed theoretically in detail in Ref. 3, fluctuations at both short (atomic) and long (meso-scopic) length-scales play a crucial role. Usually, short wavelength fluctuations exist only around $T_c$. In the present special case, however, frustrated antiferromagnetic correlations between Mn spins become enhanced at low temperature and, via the ferromagnetic coupling between Mn and Co, short length-scale fluctuations of the order several Å can be induced in the Co spin system with a long helical periodicity of 100 nm order (*19-22*). Thus, two kinds of fluctuations with a similar short length scale are found to promote topological phase stability in $Co_7Zn_7Mn_6$; one is the thermal fluctuations against helical order that stabilize the conventional SkX phase just below $T_c$, and the other is frustration-induced fluctuations that stabilize the novel DSk phase at low temperatures just above $T_g$. With this scenario, the absence of the DSk phase below $T_g$, where dynamical fluctuations are suppressed due to spin freezing, is reasonably understood. Importantly, the skyrmion size here is governed by the DMI, this being much larger than the length-scale of frustrated antiferromagnetic spin modulation. This makes the disordered skyrmions observed here distinct from the atomic length-scale skyrmion states expected in theory for genuinely frustrated systems with centrosymmetric lattices (*11-13*). Thus, the present study demonstrates a novel mechanism of skyrmion formation in DMI-based helical magnet that exploits frustration to stabilize topological non-coplanar spin structures.

## Materials and Methods

**Sample preparation:** Polycrystalline samples of $(Co_{0.5}Zn_{0.5})_{20-x}Mn_x$ ($0 \leq x \leq 20$) were synthesized from pure Co, Zn and Mn metals with nominal concentrations. These metals were sealed in evacuated quartz tubes and heated above 1000°C, cooled down to 925°C at the rate of 1°C min$^{-1}$, annealed for 2-4 days and finally quenched to water. Phase purity with a $\beta$-Mn-type crystal structure was confirmed using powder X-ray diffraction for each sample. Single-crystalline bulk samples of $Co_7Zn_7Mn_6$ were grown by the Bridgman method. The crystal orientation was determined using the X-ray diffraction Laue method (Fig. S1C) and the samples were cut along the (100), (010) and (001) planes with rectangular shapes for small angle neutron scattering (SANS) and ac susceptibility measurements as shown in Figs. S1A and B, respectively. Due to the difference in shape between the samples used in the ac susceptibility and the SANS measurements, their demagnetization factors are different. In order to correct for this difference, the magnetic field ($H$) values for the ac susceptibility measurements are calibrated to be $H_c = 2.7*H$. For all the figures related to ac susceptibility measurements in the main text and Supplementary Materials, the calibrated value $H_c$ is used with the notation of $H$.

For the Lorentz transmission electron microscopy (LTEM) measurement, a tiny thin-plate sample with (001) face and thickness of approximately 150 nm was prepared from a single-crystalline bulk sample using a focused ion beam (FIB) of Ga.

**Magnetization measurement:** (Dc) magnetization measurements for polycrystalline samples of $(Co_{0.5}Zn_{0.5})_{20-x}Mn_x$ were performed using the vibrating sample magnetometer (VSM) mode of a superconducting quantum interference device magnetometer (MPMS3, Quantum Design).

**Ac susceptibility measurement:** Ac susceptibility measurements for a single-crystalline sample of $Co_7Zn_7Mn_6$ were performed using the ac susceptibility measurement mode of the MPMS3. Both the static magnetic field and the ac excitation field (1 Oe) were applied along a [100] direction. The ac frequency $f$ was fixed to be 193 Hz except for the frequency dependent measurements.

**SANS measurement:** SANS measurements on a single-crystalline sample of $Co_7Zn_7Mn_6$ were performed using the instrument D33 at the Institut Laue-Langevin (ILL), Grenoble, France. Neutrons with a wavelength of 10 Å were collimated over 12.8 m before the sample. The scattered neutrons were counted by a 2D position-sensitive multi-detector located 13.4 m behind the sample. The mounted single-crystalline sample was installed into a horizontal field cryomagnet so that the field direction was parallel to the [001] direction. Maintaining the $H \parallel [001]$ geometry, the cryomagnet was rotated (rocked) around the vertical [010] direction in the range from −5° to 5° (from −6° to 12° only for $H \parallel$ beam geometry in the process of ZFC, and field scans at 146 K, 130 K and 100 K), and tilted around the horizontal [100] (or [001]) direction in the range from −5° to 5°. These two scans were performed for both $H \parallel$ beam and $H \perp$ beam configurations. All the SANS images shown in the main text and Supplementary Materials are obtained by summing over the above rotation and tilt rocking scans, and the scales of horizontal axis ($q_x$) and vertical axis ($q_y$) are fixed to be $-0.14 \text{ nm}^{-1} \leq q_x \leq 0.14 \text{ nm}^{-1}$ and $-0.14 \text{ nm}^{-1} \leq q_y \leq 0.14 \text{ nm}^{-1}$.

**LTEM measurement:** LTEM measurements for a single-crystalline thin-plate sample of Co$_7$Zn$_7$Mn$_6$ were performed with a transmission electron microscope (JEM-2800). A magnetic field was applied perpendicular to the (001) plate and its magnitude was controlled by tuning the electric current of the objective lens. Due to the thin-plate shape (thickness ~ 150 nm), its demagnetization factor and thus magnetic field scale are different from those of the bulk samples. For Fig. 6D, we extracted an in-plane magnetization image with using a software package QPt based on the transport-of-intensity equation:

$$\frac{2\pi}{\lambda} \frac{\partial I\,(xyz)}{\partial z} = \nabla_{xy}\left[I\,(xyz)\nabla_{xy}\varphi(xyz)\right] \tag{E1}$$

Here, $I\,(xyz)$ and $\varphi(xyz)$ represent the intensity and the phase of the electron beam, respectively. The analysis of electron beam intensity based on the under-focused and over-focused LTEM images enables to obtain the phase image $\varphi(xyz)$. According to Maxwell-

Ampère equations, $\nabla_{xy}\varphi(xyz)$ is related to the in-plane component of the magnetization via

$$\nabla_{xy}\varphi(xyz) = -\frac{e}{\hbar}(\boldsymbol{M} \times \boldsymbol{n})t, \tag{E2}$$

where $\boldsymbol{M}$, $t$, and $\boldsymbol{n}$ are the magnetization, the sample thickness, and the unit vector perpendicular to the sample surface, respectively.

**Acknowledgments**

**General**: We are grateful to N. Nagaosa, W. Koshibae, F. Kagawa and T. Nakajima for fruitful discussions.

**Funding:** This work was supported by JSPS Grant-in-Aids for Scientific Research (S) No.24224009 and for Young Scientists (B) No.17K18355, the Swiss National Science Foundation (SNF) Sinergia network 'NanoSkyrmionics' (Grant No. CRSII5-171003), the SNF projects 153451 and 166298, and the European Research Council project CONQUEST.

**Author contributions:** Y. Taguchi, Y. Tokura and H.M.R. jointly conceived the project. A.K., K.K. and Y. Tokunaga prepared the bulk samples. K.K. performed the magnetization and ac susceptibility measurements. J.S.W., K.K., C.D.D. and expected at the center of each pattern is masked out. (i) The helical (H) state forms three domains with single-$q$ || [100], [010] or [001], respectively, resulting in 4 spots in both the geometries. (ii) In the conical (C) state with $q$ || $H$ || [001], only 2 spots are observed in the $H \perp$ beam geometry. (iii) The triangular-lattice skyrmion crystal (SkX) state forms two domains with one of the triple-$q$ || [100] or [010] (degenerate preferred $q$-directions), respectively, resulting in 12 spots in the $H$ || beam geometry, and 2 spots in the $H \perp$ beam geometry. (iv) In the disordered helical (DH) state, the spots are broadened compared with (i). (v) In the three-dimensionally disordered skyrmion (DSk) state, a spherical $q$ distribution manifests itself as a ring in both geometries. (**C**) SANS images observed in $Co_7Zn_7Mn_6$ at 146 K and at 0 T and 0.03 T in the $H$ || beam and $H \perp$ beam geometries, respectively. The 0 T and 0.03 T patterns represent (i) helical and (iii) SkX (plus (ii) conical) states, respectively. 12 spots in $H$ || beam geometry indicate that the SkX state indeed consists of two kinds of domains with one of the triple-$q$ || [100] or [010]. Note that the intensity scale of the color plots varies between each panel.

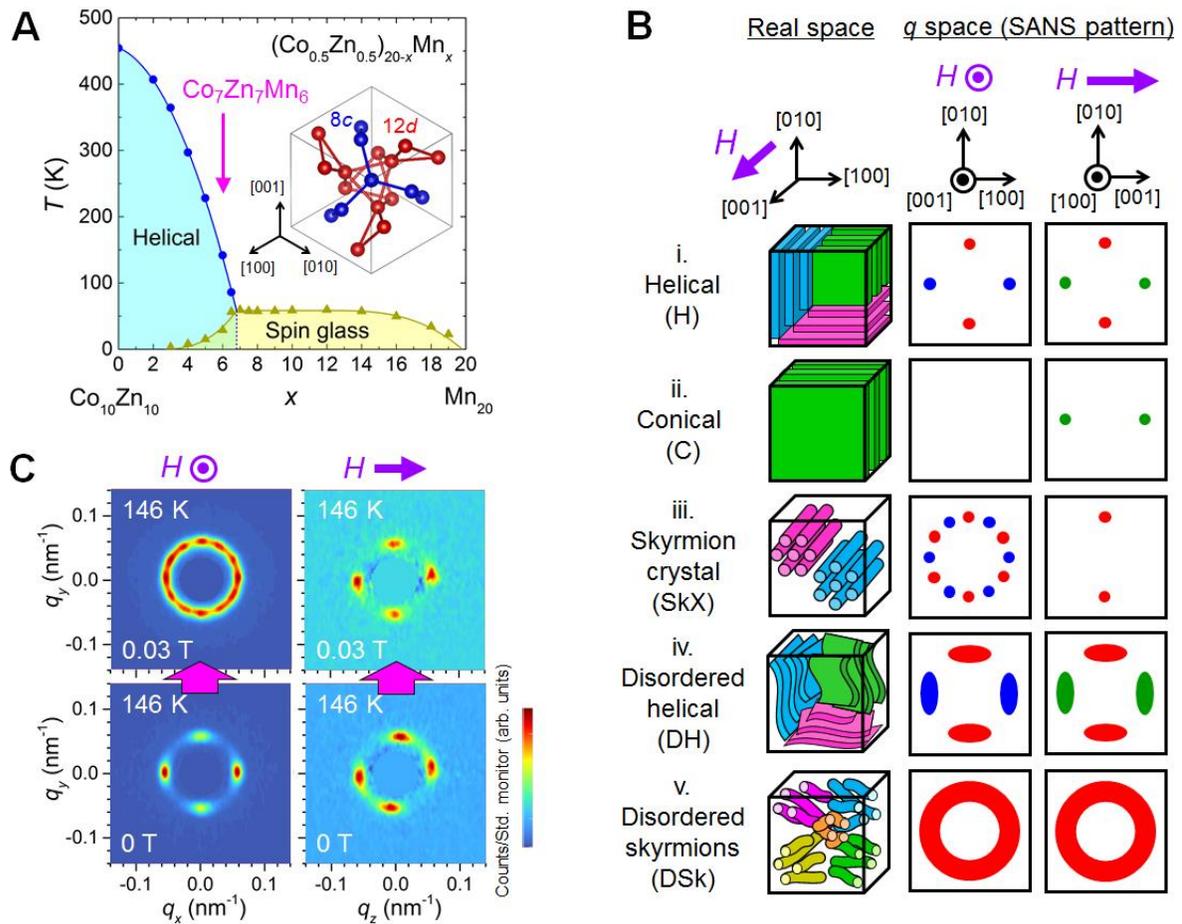

**Fig. 1. Temperature - Mn concentration phase diagram of $(Co_{0.5}Zn_{0.5})_{20-x}Mn_x$, schematic small angle neutron scattering (SANS) patterns and SANS images at 146 K in $Co_7Zn_7Mn_6$.** (**A**) The zero-field magnetic phase diagram of $(Co_{0.5}Zn_{0.5})_{20-x}Mn_x$ ($0 \leq x \leq 20$) in the $T$ (temperature) vs. $x$ (Mn composition) plane, as determined by magnetization measurements (see Fig. S2A). The inset shows a schematic of a $\beta$-Mn-type structure (space group: $P4_132$) as viewed along the [111] axis. (**B**) Schematic of magnetic structures in real space and the corresponding small angle neutron scatting (SANS) patterns in the $H$ (magnetic field) ∥ neutron beam and $H \perp$ neutron beam geometries. The direct-beam spot

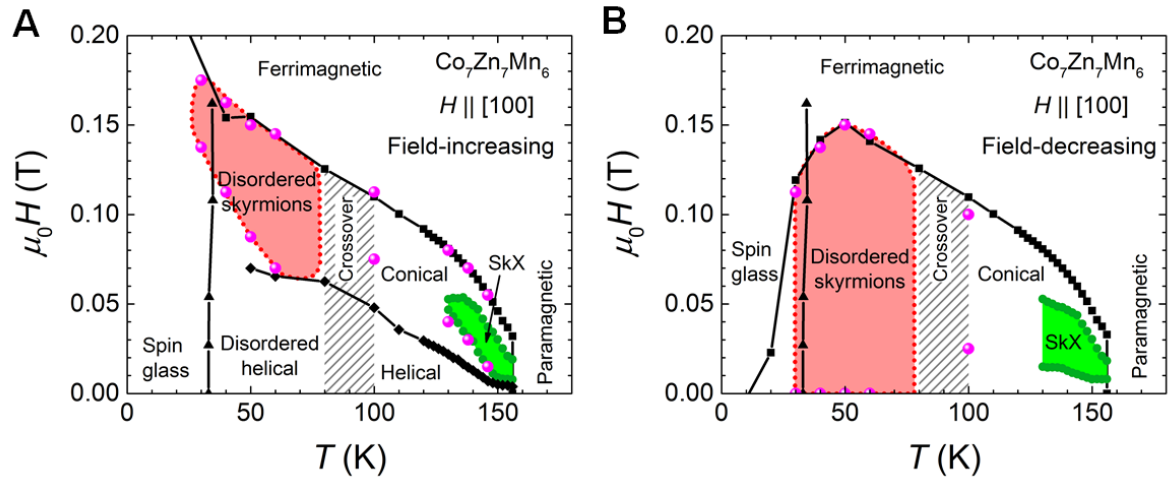

**Fig. 2. Temperature - field phase diagrams in $Co_7Zn_7Mn_6$.** $T$-$H$ phase diagrams in $Co_7Zn_7Mn_6$ determined by ac susceptibility and SANS measurements (see Figs. S3B, S4D,E and S8) (**A**) in the field-increasing runs after zero-field cooling (ZFC) and (**B**) in the field-decreasing runs from the induced ferrimagnetic (higher $H$) region. The helical-conical and conical-ferrimagnetic phase boundaries are indicated by black diamonds and squares, respectively. The phase boundaries of the conventional skyrmion crystal (SkX) phase are indicated by green circles. The crossover region around 90 K, below which the helical state becomes disordered, is indicated by gray hatching. Spin glass transition temperatures around 30 K are indicated by black triangles. Pink circles show the magnetic fields where either 12 spot or ring-like SANS patterns are observed. The equilibrium phase (plus metastable state in the case of field-decreasing processes) of disordered skyrmions is indicated by a red color region. Note that the SkX and conical states, and conical and disordered skyrmion states coexist in broad regions, and only the majority phase is indicated in the phase diagrams.

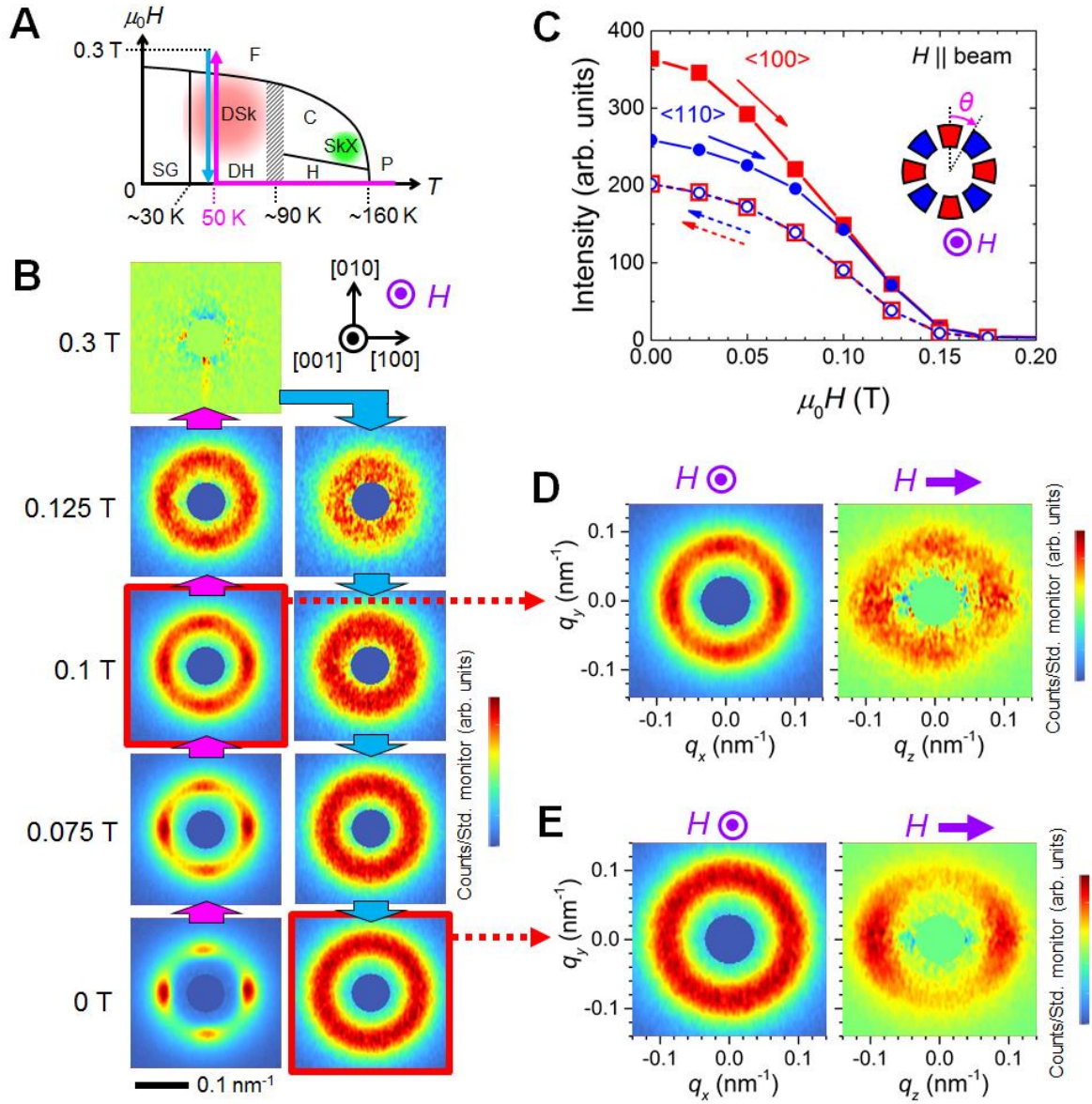

**Fig. 3. Field dependence of the SANS patterns at 50 K in Co$_7$Zn$_7$Mn$_6$.** (A) Schematic of the measurement process. The field-increasing run from 0 T to 0.3 T after ZFC and the subsequent field-decreasing run from 0.3 T to 0 T are denoted by pink and light blue arrows, respectively. In the schematic phase diagram, we use the following notations; P:

paramagnetic, H: helical, C: conical, SkX: skyrmion crystal (green region), F: ferrimagnetic, DH: disordered helical, DSk: disordered skyrmions (red region), and SG: spin glass. (**B**) SANS images at selected fields in the $H \parallel$ beam geometry. Note that the intensity scale of the color plots varies between each panel. (**C**) Field dependence of integrated SANS intensities. The intensities for directions close to <100> over the regions at $\theta = 0 \pm 15°$, $90 \pm 15°$, $180 \pm 15°$, $270 \pm 15°$ (red region in the inset) are indicated by red squares. The intensities for directions close to <110> over the regions at $\theta = 45 \pm 15°$, $135 \pm 15°$, $225 \pm 15°$, $315 \pm 15°$ (blue region in the inset) are indicated by blue circles. Here, $\theta$ is defined as the clockwise azimuthal angle from the vertical [010] direction. The field-increasing and field-decreasing runs are indicated by closed and open symbols with the same colors, respectively. (**D, E**) SANS images (D) at 0.1 T in the field-increasing run and (E) at 0 T after the field-decreasing run in both geometries, respectively. In (D) and (E), the data for $H \perp$ beam geometry were collected within the same field-sweeping process (shown schematically in (A)) as used for collecting the data shown in (B). Note that the intensity scale of the color plots varies between each panel.

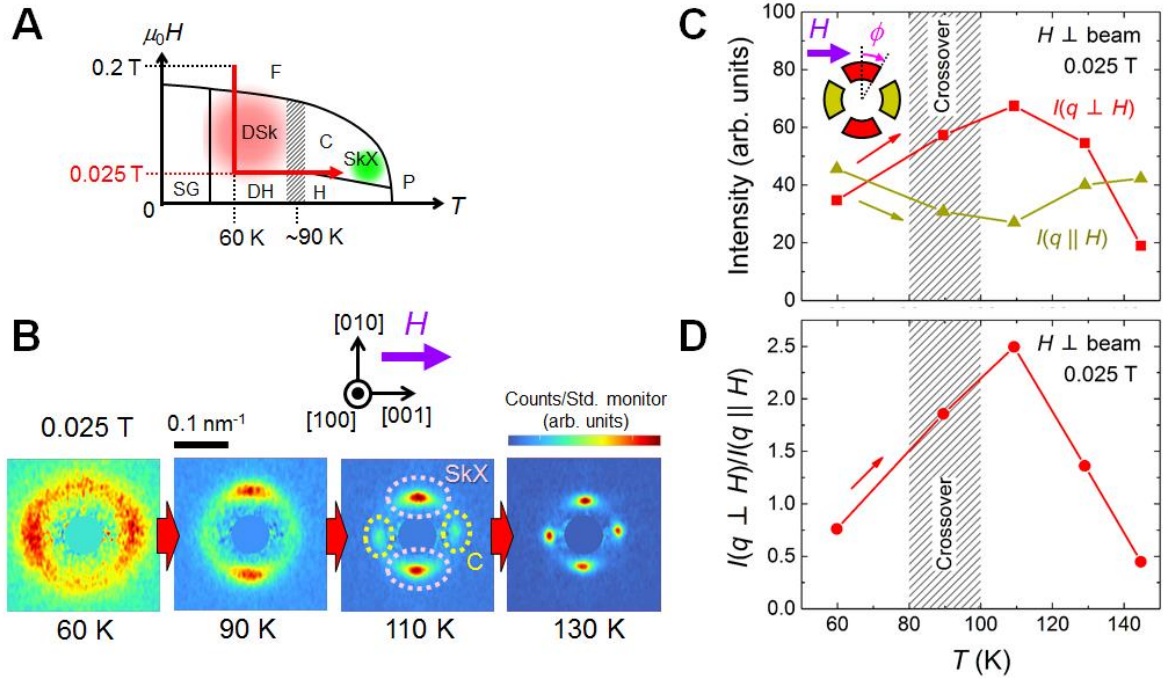

**Fig. 4. SANS patterns during the field warming process in $Co_7Zn_7Mn_6$.** (**A**) Schematic of the measurement process. (**B**) SANS images at selected temperatures in the $H \perp$ beam geometry for the field-warming (FW) process at 0.025 T after a field decrease at 60 K. The intensity scale of the color plots varies between each panel. It is noted that the SANS images both for this FW process and the zero-field warming (ZFW) process (Fig. S9C) were obtained from two rocking scans covering the same scanning region where the cryomagnet is rotated from −5° to 5° around both the vertical and horizontal axes. Pink and yellow dotted circles highlight the SANS intensities allocated to the SkX state and conical state, respectively. (**C**) Temperature dependence of integrated SANS intensities. The intensities integrated over the regions nearly perpendicular to the field at $\phi = 0 \pm 30°$, $180 \pm 30°$ (red region in the inset) are indicated by red squares [$I(q \perp H)$]. The intensities

integrated over the regions nearly parallel to the field at $\phi = 90 \pm 30°$, $270 \pm 30°$ (yellow region in the inset) are indicated by yellow triangles [$I(q\|H)$]. Here, $\phi$ is defined as the clockwise azimuthal angle from the vertical [010] direction. (**D**) Temperature dependence of the SANS intensity ratio, $I(q\perp H)/I(q \| H)$. The intensity ratio becomes smaller than 1 at 146 K because the volume fraction of a conical state is larger than that of an equilibrium SkX state (see also Fig. S4). For panels C and D, the crossover region, above (below) which both helical and skyrmion states are ordered (disordered), is indicated with gray hatching.

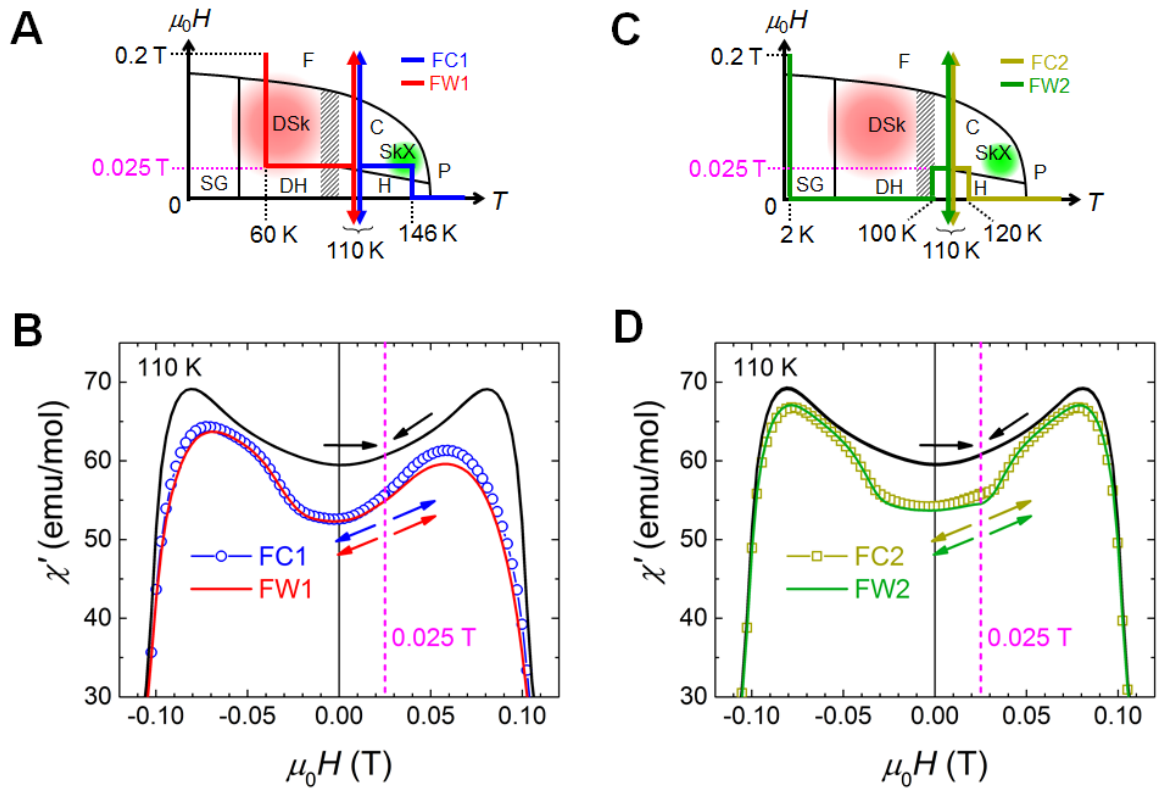

**Fig. 5. Field-swept ac susceptibility after several cooling/warming processes in Co$_7$Zn$_7$Mn$_6$.** (**A**) Schematic of measurement processes for panel B. FC1 (blue line) is a field-cooling (FC) process passing through the SkX phase; (i) ZFC to 146 K, (ii) field increase to 0.025 T and (iii) FC to 110 K. FW1 (red line) is a field-warming (FW) process passing through the DSk phase; (i) ZFC to 60 K, (ii) field increase to 0.2 T, (iii) field decrease to 0.025 T and (iv) FW to 110 K. (**B**) Field dependence of real-part of ac susceptibility ($\chi'$) at 110 K after FC1 (blue circles) and FW1 (red line). Black lines show the field-returning runs from high-field ferrimagnetic regions. (**C**) Schematic of measurement processes for panel D. FC2 (yellow line) is a FC process bypassing the SkX phase; (i) ZFC to 120 K, (ii) field increase to 0.025 T and (iii) FC to 110 K. FW2 (green line) is a FW process bypassing the DSk phase; (i) ZFC to 2 K, (ii) field increase to 0.2 T, (iii) field decrease to 0 T, (iv) ZFW to 100 K, (v) field increase to 0.025 T and (vi) FW to 110 K. (**D**) Field dependence of $\chi'$ at 110 K after FC2 (yellow squares) and FW2 (green line). Black lines show the field-returning runs from high-field ferrimagnetic regions. In panels B and D, the black lines for FC1 and FW1, and for FC2 and FW2, completely overlap with each other.

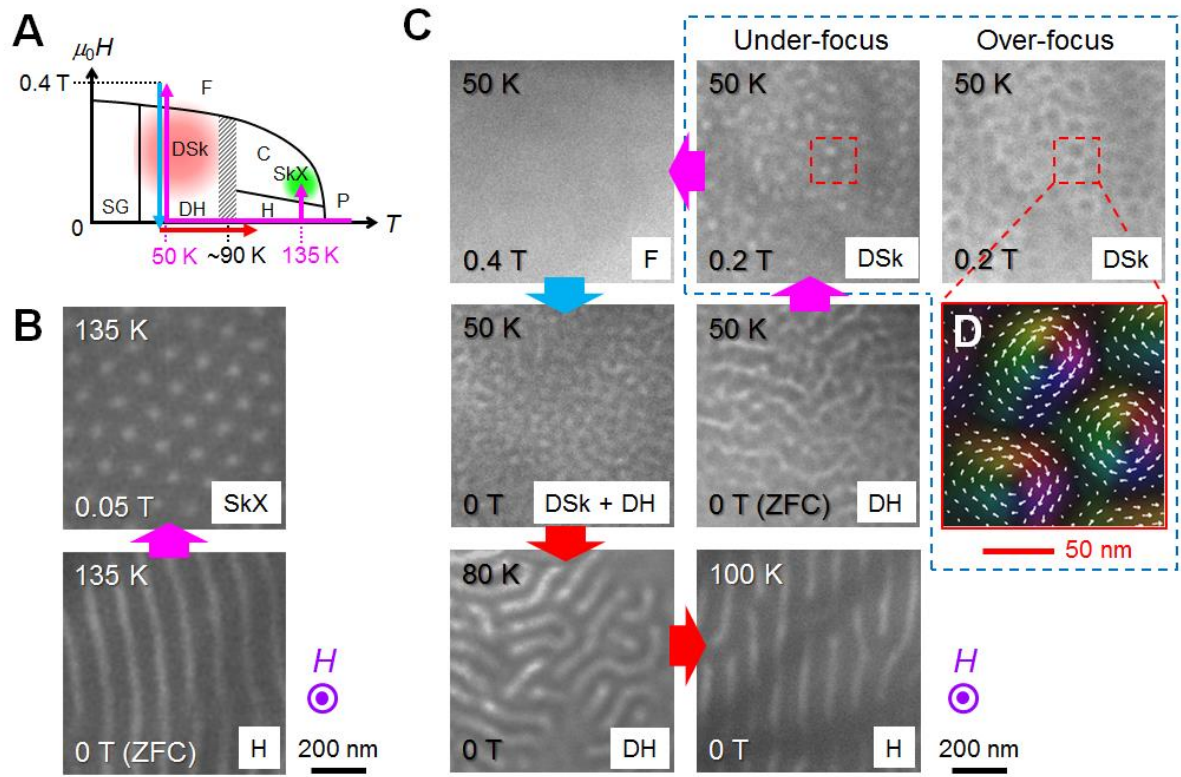

**Fig. 6. Lorentz transmission electron microscopy (LTEM) measurements on a (001) thin-plate sample of $Co_7Zn_7Mn_6$.** (**A**) Schematic illustration of the measurement processes. The field-increasing runs at 135 K and 50 K after ZFC from room temperature are denoted by pink arrows. The field-decreasing run from 0.4 T to 0 T at 50 K and the subsequent zero-field warming (ZFW) process are denoted by light blue and red arrows, respectively. (**B**) Under-focused LTEM images at 135 K and at 0 T and 0.05 T. (**C**) Under-focused LTEM images at selected fields at 50 K and at selected temperatures in the subsequent ZFW process. Only for the image at 50 K and 0.2 T, the corresponding over-focused image is also shown at right side. The assignment of the LTEM images on each panel is given such as H, SkX, DH, DSk, F, and DSk + DH (mixed state). (**D**) Color coding of in-plane

magnetization (white arrows) deduced from a transport-of-intensity equation analysis for the areas marked with the red dashed square in the under-focused and over-focused images at 50 K and 0.02 T.

Supplementary Materials for

# Disordered skyrmion phase stabilized by magnetic frustration in a chiral magnet

K. Karube[1]*†, J. S. White[2]†, D. Morikawa[1], C. D. Dewhurst[3], R. Cubitt[3], A. Kikkawa[1], X. Z. Yu[1], Y. Tokunaga[4], T. Arima[1,4], H. M. Rønnow[5], Y. Tokura[1,6], Y. Taguchi[1]

[1]RIKEN Center for Emergent Matter Science (CEMS), Wako 351-0198, Japan.

[2]Laboratory for Neutron Scattering and Imaging (LNS), Paul Scherrer Institute (PSI),

CH-5232 Villigen, Switzerland.

[3]Institut Laue-Langevin (ILL), 71 avenue des Martyrs, CS 20156, 38042 Grenoble cedex 9,

France

[4]Department of Advanced Materials Science, University of Tokyo, Kashiwa 277-8561,

Japan.

[5]Laboratory for Quantum Magnetism (LQM), Institute of Physics,

École Polytechnique Fédérale de Lausanne (EPFL), CH-1015 Lausanne, Switzerland.

[6]Department of Applied Physics, University of Tokyo, Bunkyo-ku 113-8656, Japan.

*Corresponding author. Email: kosuke.karube@riken.jp

†These authors contributed equally to this work

**Magnetic properties in $(Co_{0.5}Zn_{0.5})_{20-x}Mn_x$**

Figure S2A shows the temperature dependence of the magnetization for $(Co_{0.5}Zn_{0.5})_{20-x}Mn_x$ for several Mn concentration ($x$)-values. The helimagnetic transition temperature $T_c$ decreases with increasing $x$. For $3 \leq x \leq 6.5$, while the helimagnetic transition is still observed, the magnetization in the field warming (FW) run after the zero-field cooling (ZFC) process shows a drop at low temperature which is not observed in the field cooling (FC) process. For $x \geq 7$, the magnetization values are much smaller than those for $x \leq 6.5$ due to the suppression of helimagnetic state with dominant ferromagnetic interaction. Instead, the typical temperature-dependences for a spin glass, i.e. a cusp and a hysteresis between FC and FW processes, are observed at ~ 60 K. In view of the systematic change of the magnetization curves across $x = 6 \sim 7$, the drop of magnetization in the FW process observed in $3 \leq x \leq 6.5$ is interpreted as due to a reentrant spin glass transition after a helimagnetic ordering. The spin glass transition temperature $T_g$ is independent of $x$ in the range of $7 \leq x \leq 14$ while the magnetization value becomes smaller with increasing $x$. For $x \geq 16$, $T_g$ smoothly decreases and finally no magnetic transition is observed at $x = 20$.

These results are summarized in the $T$ (temperature) - $x$ phase diagram of Fig. 1A in the main text and Fig. S2B. From the $Mn_{20}$ ($\beta$-Mn) side, the spin glass phase is induced by the substitution of $Co_{0.5}Zn_{0.5}$ by an amount of just 5%, as reported in other $\beta$-Mn alloys partially substituted with, for example, Al (*19*) and Co (*21*). The spin glass phase exists over a wide $x$ range and coexists with the helical phase for $3 \leq x \leq 6.5$. This demonstrates the existence of Mn-Mn antiferromagnetic correlations even in the helical phase. A reentrant spin glass transition after a long-range magnetic ordering as in the present case has been reported in other materials, including both ferromagnetic (*23*) and antiferromagnetic (*24*) systems, and in all of which the reentrant spin glass state is considered to microscopically coexist with the long-range ordered state.

As shown in Fig. S2C, the magnetization at 2 K and 7 T is maximized around $x = 2$. This suggests that Co and Mn are essentially ferromagnetically coupled at least in the low $x$ region while antiferromagnetic Mn-Mn interactions become increasingly significant for higher $x$ values. In Fig. S2D, we plot the lattice constant of the $\beta$-Mn-type structure as a function of $x$. Although the lattice constants of $Co_{10}Zn_{10}$ and $Mn_{20}$ are very similar (~ 6.32 Å), those of the alloys in between show a broad maximum at $x \sim 12$. This may reflect changes in the magnetic state at the atomic level as the Mn substitution proceeds.

**Zero-field cooling process in $Co_7Zn_7Mn_6$**

Figure S3 shows ac susceptibility and SANS measurements obtained during the ZFC process.

As shown in Fig. S3B, the real part of the ac susceptibility $\chi'$ shows a large increase around 160 K due to the helimagnetic transition, gradually decreasing below 90 K, and finally dropping more sharply around 30 K, where the imaginary part $\chi''$ shows a large peak. The temperature dependence of $\chi'$ at several ac frequencies $f$ is shown in Fig. S3C. The temperature $T(f)$ at the drop of $\chi'$, determined as the inflection point, increases with increasing $f$. This frequency dependence is typical for a spin glass and is well fitted to a power law, $f^{-1} = \tau_0 [T(f)/T_0 - 1]^{-zv}$. Here, $T_0$ is the spin freezing temperature in the limit of $f$

→ 0, and $\tau_0$ is a relaxation time of spin flop at $T = 2T_0$, and $zv$ are critical exponents. The obtained value of $T_0 = 24.6$ K is reasonable being close to the value of $T_g = 25.8$ K determined by the inflection point of the dc magnetization under 20 Oe. The obtained value of $\tau_0 = 5.3 \times 10^{-8}$ s is much longer than typical values for spin glasses ($\sim 10^{-13}$ s), indicating a microscopic coexistence of the spin glass and helical state. Similar values of $\tau_0 \sim 10^{-8}$ s have been reported in other materials exhibiting a coexistence of spin glass and ferromagnetism (*23*).

As shown in Fig. S3D, the spots in the SANS pattern broaden below ~ 100 K and are accompanied by a clear increase in $q$, indicating that the helical state becomes partially disordered and that the periodicity decreases below 100 K. To discuss these temperature variations of the SANS patterns quantitatively, the integrated intensities for directions close to <100> and <110> are plotted as a function of temperature in Fig. S3F. The integrated intensities for both directions decreases and the azimuthal angle-dependent $q$ distribution becomes more uniform below ~ 90 K. In Fig. S3E, line profiles of the azimuthal angle-averaged intensity are plotted against $q$ and fitted to Voigt functions. The center $q_0$ and the full-width-half-maximum (FWHM) of the Voigt functions, and also their inverses (the helical periodicity $\lambda$ and the correlation length $\xi$) are plotted as a function of temperature in Figs. S3G and H, respectively. The intensity profile is observed to broaden along the radial $q$ direction below ~ 90 K, and the helical periodicity shrinks from 110 nm to 73 nm. These quantities saturate and become almost temperature-independent below 60 K.

The disordering of the helical state below 90 K, and the subsequent spin glass transition around 30 K in the present case are readily attributed to an enhanced influence of frustrated antiferromagnetic Mn-Mn interactions.

**Field-sweeping processes in Co$_7$Zn$_7$Mn$_6$**

In Figs. S4-S8, we show SANS and ac susceptibility measurements obtained during field-sweeping processes at several temperatures.

Figure S4 shows the results at 146 K and 130 K. In the ac susceptibility measurement at 146 K (Fig. S4D), $\chi'$ shows a clear dip structure over a region of 0.015-0.04 T due to the formation of the SkX. In the SANS measurement at 146 K (Fig. S4B, left), 12 spot patterns (two-domain triangular-lattice SkX) are observed in the $H \parallel$ beam geometry under magnetic fields. The integrated intensities for directions close to <100> and <110> overlap with each other for 0.015-0.05 T and show a large peak due to the formation of SkX (Fig. S4F). In the $H \perp$ beam geometry (Fig. S4B, right), in addition to the 2 vertical spots (SkX), 2 horizontal spots (conical) are observed. The integrated intensities observed in the region parallel to the field is stronger than that perpendicular to the field for 0.01-0.02 T, and becomes slightly weaker for 0.03-0.04 T due to the change in the volume fraction between SkX state and conical state (Fig. S4H).

In the ac susceptibility measurement at 130K (Fig. S4E), the anomaly in $\chi'$ due to SkX formation is barely observed for 0.04-0.06 T. In the SANS measurement at 130K (Fig. S4C, left), ring patterns (orientationally-disordered triangular-lattice SkX) are observed in the $H$

∥ beam geometry for 0.04-0.08 T, where the total integrated intensities for directions close to <100> and <110> overlap with each other, forming a small peak (Fig. S4G). In the $H \perp$ beam geometry (Fig. S4C, right), however, the 2 vertical spots (SkX) are much weaker than the 2 horizontal spots (conical) under magnetic field, as also indicated in Fig. S4I. Therefore, it is safely concluded that 130 K is the low temperature limit of the equilibrium SkX phase.

Figure S5 shows the results at 100 K. In the ac susceptibility measurement (Fig. S5C), the dip structure of $\chi'$ is not discerned any more. In the SANS measurement (Fig. S5B, D), ring-like patterns with very weak intensities are barely observed in the $H \parallel$ beam geometry for 0.075-0.1 T. The SANS intensity in the field-decreasing run is even weaker than that in the field-increasing run. These results indicate that the volume fraction of any skyrmion state is tiny and that the conical state is dominant at 100 K.

Figure S6 shows the results at 50 K. In the SANS measurement in the $H \parallel$ beam geometry (Fig. S6B), the pattern with 4 broad spots (disordered helical) changes to a ring pattern above 0.1 T in the field increasing run, and the ring pattern appears again and persists to zero field in the subsequent field-decreasing run from the ferrimagnetic state. In the $H \perp$ beam geometry (Fig. S6C), elliptically deformed patterns are observed under magnetic fields and remain at zero field after the subsequent field-decreasing run. This elliptical pattern is interpreted to be due to a superposition of a ring pattern and 2 broad spots at horizontal positions. Therefore, the three-dimensional $q$ distribution corresponds to a superposition of a spherical shell and 2 broad spots, the latter of which correspond to a disordered conical state. The integrated intensity of the spherical pattern is larger than that due to the additional 2 broad spots (Fig. S6F). Similar field dependent behavior was also observed at 60 K, 40 K and 30 K. The spherical SANS pattern observed at low temperatures is consistent with either (i) three-dimensionally disordered helical state or (ii) three-dimensionally disordered skyrmions. As shown in Fig. 4 in the main text, we found that two-dimensional SkX state (multiple-$q \perp H$) is realized above 90 K as a metastable state in the subsequent field-warming (FW) processes starting from 60 K, and hence we conclude that disordered skyrmions are more plausible than disordered helical state to explain the spherical SANS pattern.

Figures S7A-C shows the results at 20 K, a temperature which is below the spin glass transition temperature $T_g \sim 30$ K. In the SANS measurement (Figs. S7B, C), the broad 4 spot pattern in the $H \parallel$ beam geometry persists to higher fields, and finally the signal disappears above 0.2 T with no transformation into a clear ring pattern. In the field-decreasing run, only scattering broadly distributed around $q = 0$ is observed below 0.1 T. These results indicate that below $T_g$, the disordered helical state persists under magnetic fields without a transition to a disordered skyrmion phase, and in the field-decreasing run, the induced ferrimagnetic state at high field persists down to near zero field.

The field dependence of the SANS intensities at all the investigated temperatures is summarized in Fig. S8. From this and the ac susceptibility results, we constructed the $T$-$x$ phase diagrams shown in Fig. 2 of the main text. There are clearly two distinct, equilibrium skyrmion phases, where the SANS intensities for directions close to <100> and <110>

overlap with each other (marked by pink triangles in Fig. S8) in the field-increasing runs. One is the conventional SkX phase stabilized at 146-130 K just below $T_c$, and the other is the novel phase of disordered skyrmions stabilized over 30-60 K above $T_g$. Once created, the disordered skyrmions in the low temperature phase are highly robust and survive down to zero field as a metastable state.

**Field warming and field cooling processes across the spin-glass transition temperature in Co$_7$Zn$_7$Mn$_6$**

Figures S7D-F show the FW process at 0.1 T from 20 K to 60 K after an initial ZFC, and the subsequent FC process from 60 K to 20 K. In the SANS measurement in the $H \parallel$ beam geometry (Figs. S7E, F), the pattern with 4 broad spots at 20 K and 0.1 T changes to a ring pattern above 40 K in the FW process. In the subsequent FC process, the ring pattern persists to 20 K. This irreversible temperature variation indicates that the equilibrium phase of disordered skyrmions is accessible from the spin glass phase and, once created, disordered skyrmions survive below $T_g$ as a metastable state. Fig. S7G shows the frequency dependence of $\chi'$ versus temperature in the FC process (0.1 T). A similar frequency dependence as for the case of the ZFC process (Fig. S3C) is observed, indicating the coexistence of spin glass and skyrmion states, for which Mn spins and Co spins are suspected to be mainly responsible, respectively. The frequency dependence was also fitted to a power law. The obtained value of $\tau_0 = 2.5 \times 10^{-8}$ s is the same order of magnitude as that obtained in the ZFC process.

**Zero-field and field warming processes after field sweepings at low temperatures in Co$_7$Zn$_7$Mn$_6$**

Figure S9 shows data obtained in the zero-field warming (ZFW) process after a field-decreasing run at 50 K. In the SANS measurement, the ring pattern changes to 4 sharp spots above 90 K in the $H \parallel$ beam geometry (Fig. S9B). The elliptical SANS intensity distribution in the $H \perp$ beam geometry also changes to one with 4 sharp spots above 90 K (Fig. S9C). After the ZFW, the temperature was reset from 130 K back to 50 K. The observed pattern at 50 K after the ZFC was not a ring (or elliptical) pattern but 4 broad spots. As discussed in the main text, this irreversible temperature variation indicates that disordered skyrmions remaining as a metastable state in zero field at 50 K changes to the equilibrium helical multi-domain state above 90 K.

Figures S10B-D show field-swept ac susceptibility $\chi'(H)$ measured after the field-decreasing process at 60 K and several subsequent FW (ZFW) processes with various warming-field $H_0$ and measurement-temperature $T_0$ conditions as illustrated in Fig. S10A. The shape of $\chi'(H)$ systematically changes with varying $T_0$ and $H_0$. An asymmetric shape of $\chi'(H)$ with respect to the sign of $H$ is a hallmark signature for the existence of metastable skyrmions: $\chi'(H)$ is largely suppressed at positive fields due to the existence of a metastable SkX state (with small value of $\chi'$ because of $q \perp H$) up to field-induced ferrimagnetic state, and therefore a suppressed volume of a conical state (with large value of $\chi'$ because of $q \parallel H$). For negative fields, the suppression of $\chi'(H)$ is relatively smaller since the metastable

SkX state is more readily destroyed in favor of helical/conical states at magnetic fields with smaller magnitude. As $H_0$ is increased, the asymmetry becomes more prominent.

For more a quantitative discussion, we define an asymmetry parameter $A \equiv (\chi'_{\max(-)} - \chi'_{\max(+)})/\chi'_{\max(-)}$. Here, $\chi'_{\max(+)}$ and $\chi'_{\max(-)}$ are maximum values of $\chi'$ at $H > 0$ and $H < 0$, respectively, as exemplified in Fig. S10B. From the above discussion, $A$ is expected to be proportional to a volume fraction of the metastable SkX state in the temperature region above the crossover where the skyrmions display improved orientational order. The $T_0$ dependence of $A$ for various $H_0$ is plotted in Fig. S10E. $A$ increases as $T_0$ is increased from 80 K to 100 K. With further increasing $T_0$ above 100 K, $A$ gradually decreases. The initial increase in $A$ corresponds to the crossover of skyrmions from a three-dimensionally disordered form (DSk) to a two-dimensionally ordered form (SkX) as a superheated metastable state, and the subsequent decrease in $A$ corresponds to the gradual destruction of the superheated metastable SkX state into the equilibrium helical/conical state. This temperature dependence shows good agreement to that of the SANS intensity ratio shown in Fig. 4D in the main text. As $H_0$ is decreased (increased), $A$ significantly becomes smaller (larger). This $H_0$ dependence of $A$ is consistent with the interpretation that $A$ is proportional to a volume fraction of the metastable SkX state because the metastable skyrmions become more fragile generally as the magnetic field is reduced.

Finally, these warming results are summarized in the $H$-$T$ state diagram shown in Fig. S10F. The temperature of the boundary (blue circles) between the superheated metastable SkX (blue region) and the equilibrium helical/conical state lowers as the field is decreased, which is in good accord with the case of metastable SkX states accessed by supercooling (*17, 26-28*).

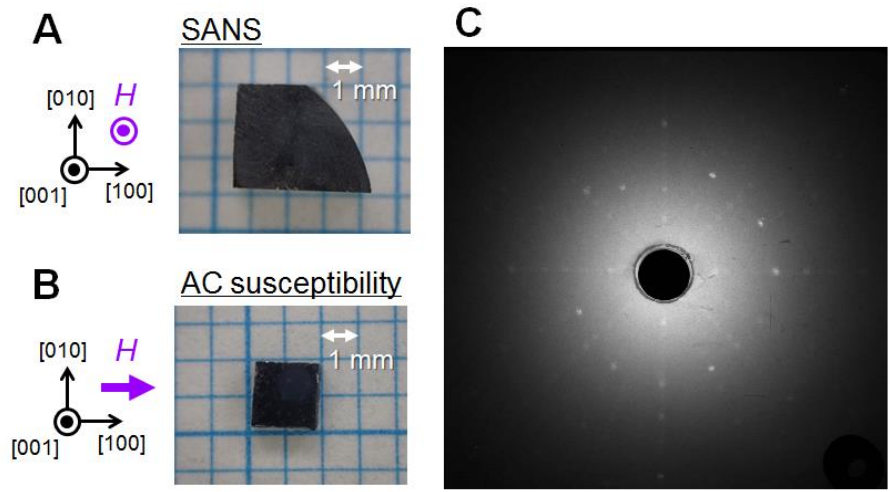

**Fig. S1. Information of single-crystalline bulk samples of Co₇Zn₇Mn₆.** Pictures of single-crystalline bulk samples of $Co_7Zn_7Mn_6$ used for (**A**) small angle neutron scattering (SANS) and (**B**) ac susceptibility and magnetization measurements. Directions of the crystal axes and applied magnetic field (*H*) are also indicated. (**C**) X-ray Laue image obtained from the (001) surface of the sample shown in panel A.

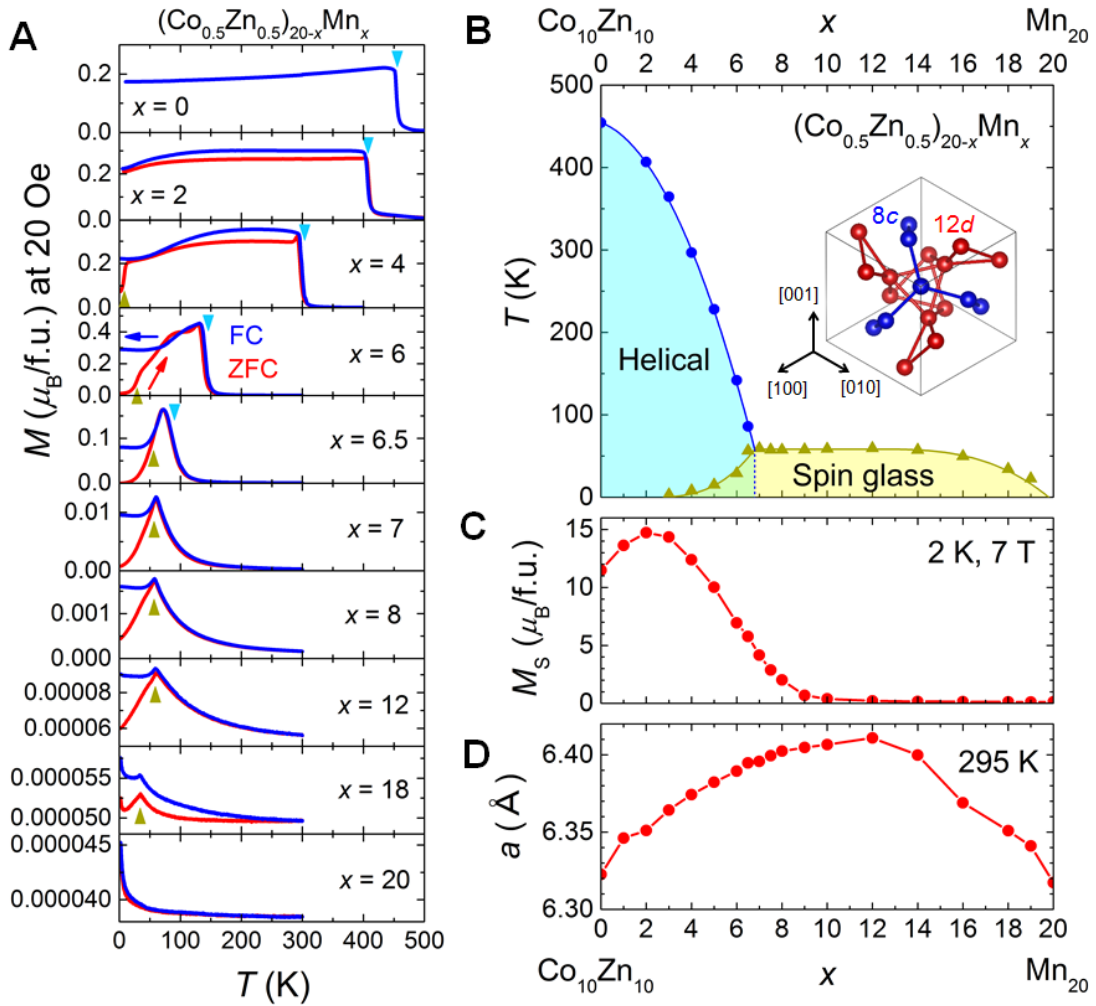

**Fig. S2. Mn concentration ($x$) dependence of polycrystalline samples of $(Co_{0.5}Zn_{0.5})_{20-x}Mn_x$.** (**A**) Temperature ($T$) dependence of magnetization ($M$) under a small magnetic field of 20 Oe. Blue and red lines show field cooling (FC), and field warming (FW) after zero-field cooling (ZFC) (just noted as "ZFC") processes, respectively. Light blue triangles, determined as the inflection point of $M(T)$, and yellow triangles, determined as the inflection point or cusp of $M(T)$ in the ZFC process, indicate the helimagnetic transition temperatures and spin glass transition temperatures, respectively. These are plotted as a function of $x$ in panel B. (**B**) $T$-$x$ phase diagram of $(Co_{0.5}Zn_{0.5})_{20-x}Mn_x$ (the same as Fig. 1A in the main text). The inset shows a schematic of the $\beta$-Mn-type structure viewed along the [111] direction. (**C**) $x$ dependence of the magnetization at 2 K and 7 T. (**D**) $x$ dependence of lattice constant $a$ of the $\beta$-Mn-type structure at room temperature.

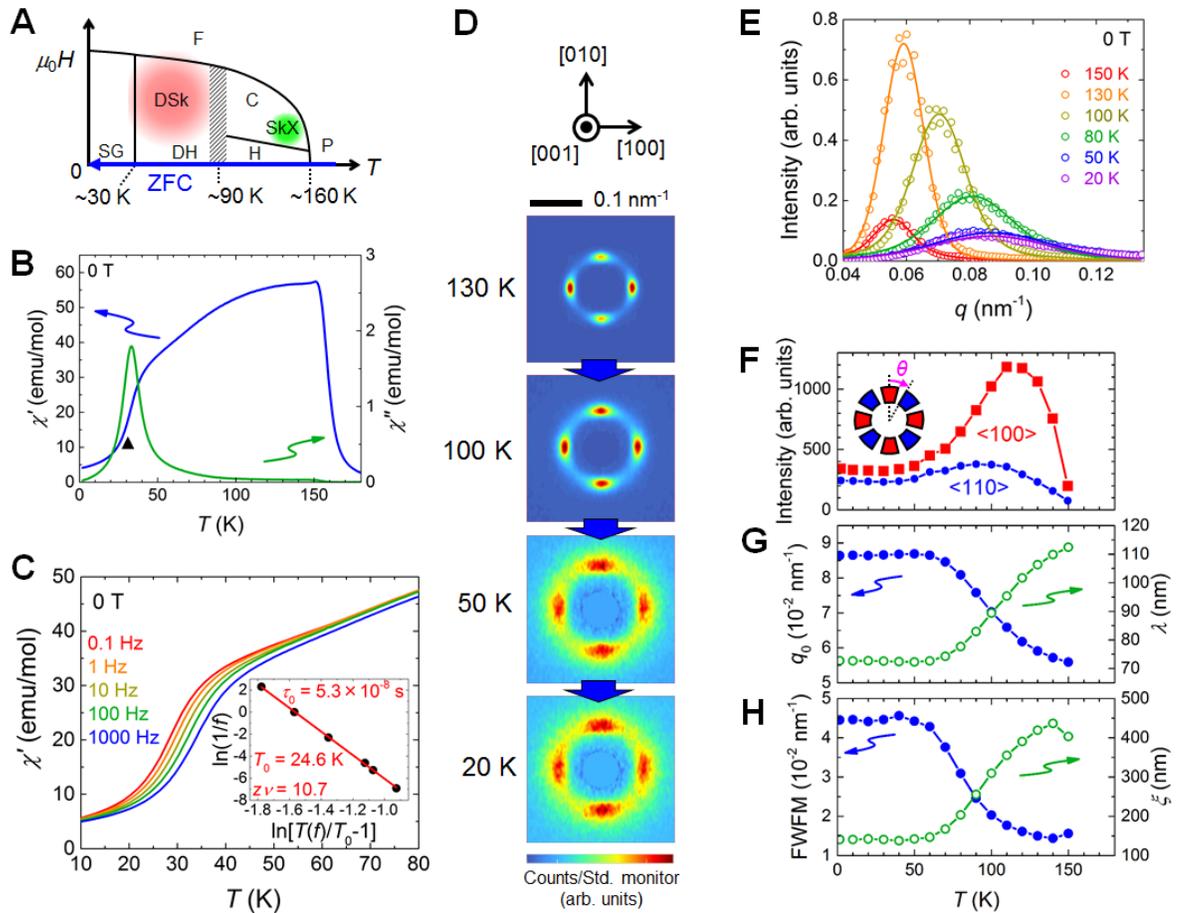

**Fig. S3. Ac susceptibility and SANS measurements in zero-field cooling (ZFC) process in $Co_7Zn_7Mn_6$.** (**A**) Schematic of the measurement process. (**B**) Temperature dependence of the real part $\chi'$ (blue line) and the imaginary part of $\chi''$ (green line) of ac susceptibility.

The spin glass transition temperature (black triangle) was determined as the inflection point of $\chi'$. This point and similar data points at different fields are plotted in the $T$-$H$ phase diagrams of Fig. 2 in the main text. (**C**) Temperature dependence of $\chi'$ at several ac frequencies $f$ from 0.1 Hz to 1000 Hz. The inset shows the $\ln(f^{-1})$ vs $\ln[T(f)/T_0-1]$ plot. $T(f)$ is defined as the inflection point associated with the drop of $\chi'$. Experimental data (black circles) are fitted to a power law (red line), $f^{-1} = \tau_0[T(f)/T_0-1]^{-zv}$. The obtained values are $\tau_0 = 5.3 \times 10^{-8}$ s, $T_0 = 24.6$ K and $zv = 10.7$, respectively. (**D**) SANS images at selected temperatures. Note that the intensity scale of the color plots varies between each panel. (**E**) Radial $q$ dependence of SANS intensities integrated over the whole azimuthal angle range at selected temperatures. Experimental data (open circles) are fitted to Voigt functions (solid lines). (**F**) Temperature dependence of SANS intensities. The intensities for directions close to <100> over the regions at $\theta = 0 \pm 15°$, $90 \pm 15°$, $180 \pm 15°$, $270 \pm 15°$ (red regions in the inset) are indicated by red squares. The intensities for directions close to <110> over the regions at $\theta = 45 \pm 15°$, $135 \pm 15°$, $225 \pm 15°$, $315 \pm 15°$ (blue regions in the inset) are indicated by blue circles. Here, $\theta$ is defined as the clockwise azimuthal angle from the vertical [010] direction. (**G**) Temperature dependence of the center $q_0$ of the Voigt function fits shown in panel E (blue closed circles) and the helical periodicity $\lambda \equiv 2\pi/q_0$ (green open circles). (**H**) Temperature dependence of the FWHM from the Voigt function fits shown in panel E (blue closed circles) and the correlation length $\xi \equiv 2\pi/\text{FWHM}$ (green open circles).

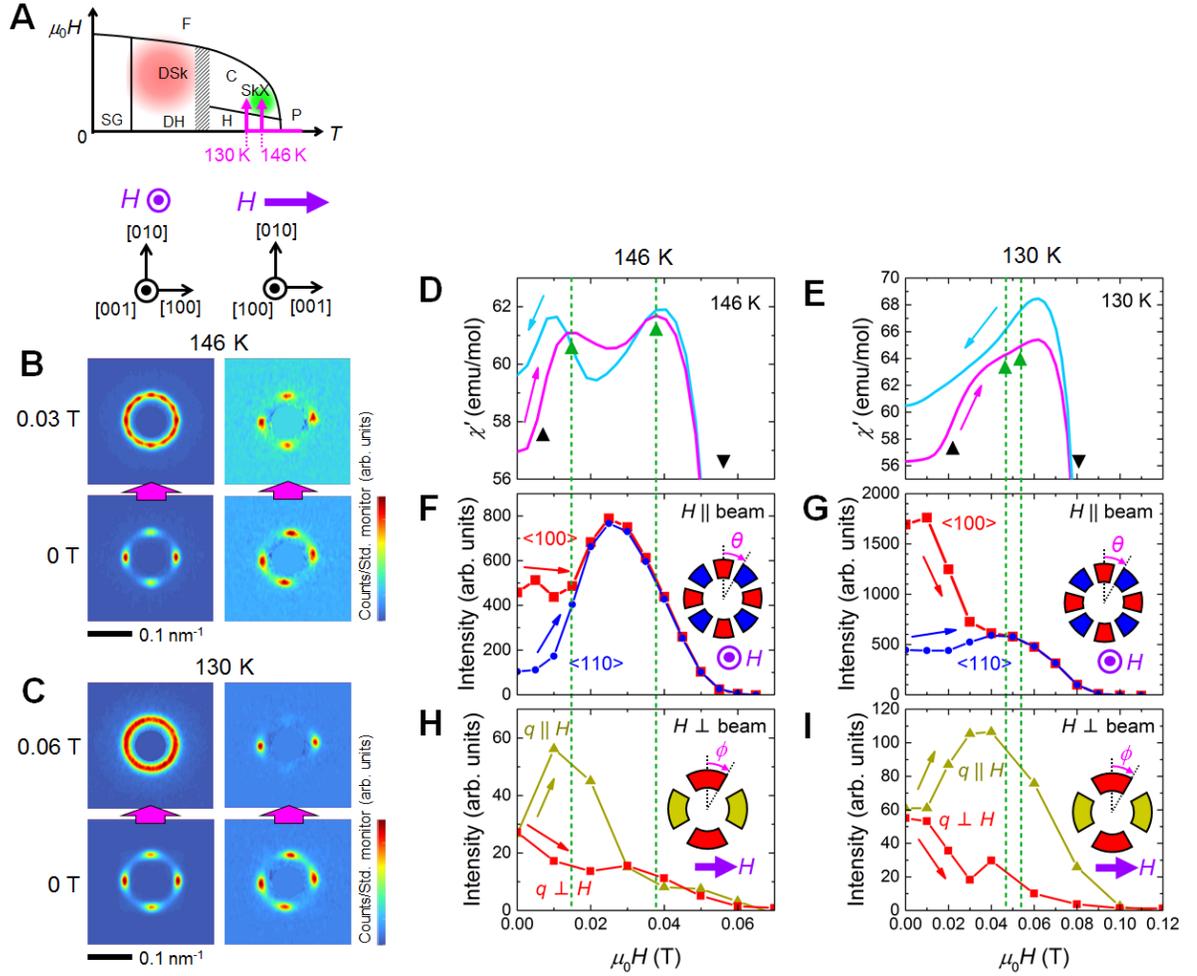

**Fig. S4. SANS and ac susceptibility measurements in the field sweeping process at 146 K and 130 K in $Co_7Zn_7Mn_6$.** (**A**) Schematic of the measurement processes. The field-increasing runs after ZFC are indicated by the pink arrows. (**B, C**) SANS images at selected fields in the $H \parallel$ beam (left) and $H \perp$ beam (right) geometries at (B) 146 K and (C) 130 K. Note that the intensity scale of the color plots varies between each panel. (**D, E**) Field dependence of $\chi'$ in the field-increasing run (pink line) and decreasing run (light blue line) at (D) 146 K and (E) 130 K. The helical-conical and conical-ferrimagnetic phase boundaries (black triangles) were determined as the inflection points of $\chi'$. The phase

boundaries (green triangles) between conical and the SkX states were determined as the peak points of $\chi'$. These boundaries and similar data points at different temperatures are plotted in the $T$-$H$ phase diagrams of Fig. 2 in the main text. (**F, G**) Field dependence of the SANS intensities in the $H \parallel$ beam geometry integrated for directions close to <100> (red squares) and <110> (blue circles) at (F) 146 K and (G) 130 K (defined similarly as in Fig. S3F). (**H, I**) Field dependence of the SANS intensities in the $H \perp$ beam geometry at (H) 146 K and (I) 130 K. The intensities integrated over the regions nearly perpendicular to the field at $\phi = 0 \pm 30°$, $180 \pm 30°$ (red region in the inset) are indicated by red squares. The intensities integrated over the regions nearly parallel to the field at $\phi = 90 \pm 30°$, $270 \pm 30°$ (yellow region in the inset) are indicated by yellow triangles. Here, $\phi$ is defined as the clockwise azimuthal angle from the vertical [010] direction.

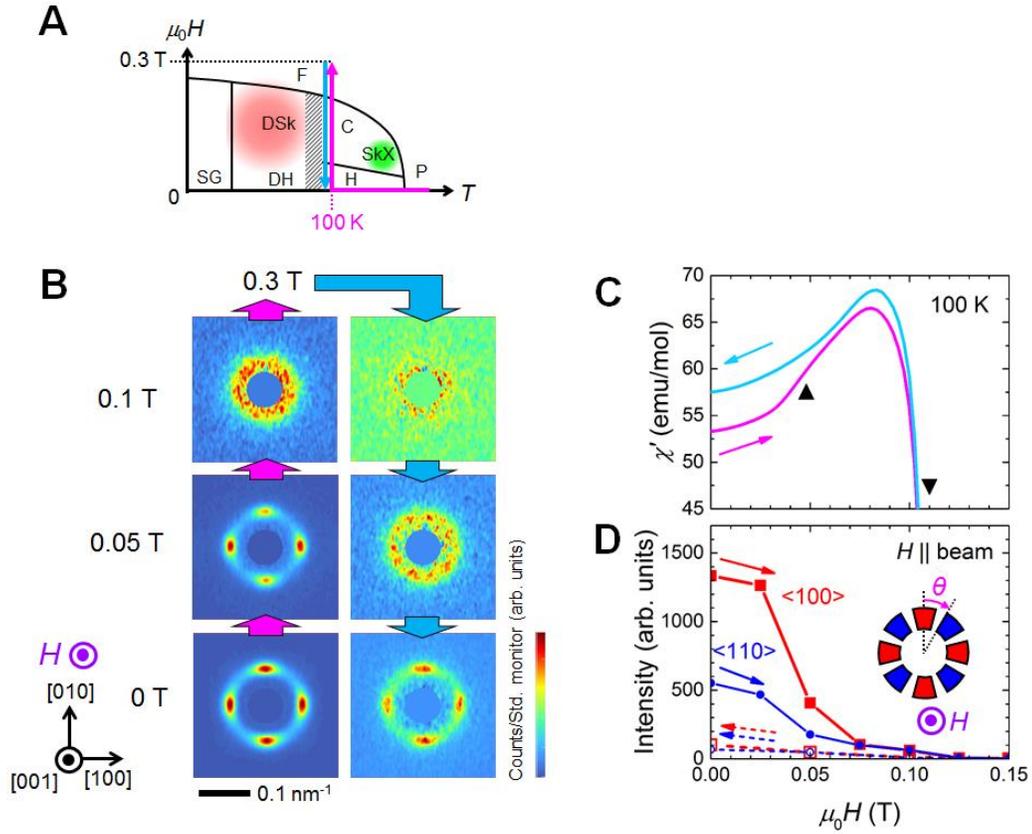

**Fig. S5. SANS and ac susceptibility measurements in the field sweeping process at 100 K in Co$_7$Zn$_7$Mn$_6$.** (**A**) Schematic of the measurement process. The field-increasing run from 0 T to 0.3 T after ZFC, and field-decreasing run from 0.3 T to 0 T are denoted by pink and light blue arrows, respectively. (**B**) SANS images at selected fields in the $H \parallel$ beam geometry. Note that the intensity scale of the color plots varies between each panel. (**C**) Field dependence of $\chi'$ in the field-increasing run (pink line) and decreasing run (light blue line). The phase boundaries indicated by triangles were determined similarly as in Figs. S4D,E. (**D**) Field dependence of the SANS intensities in the $H \parallel$ beam geometry integrated for directions close to <100> (red squares) and <110> (blue circles) (defined similarly as in

Fig. S3F). The field-increasing and field-decreasing runs are indicated by closed and open symbols with the same colors, respectively.

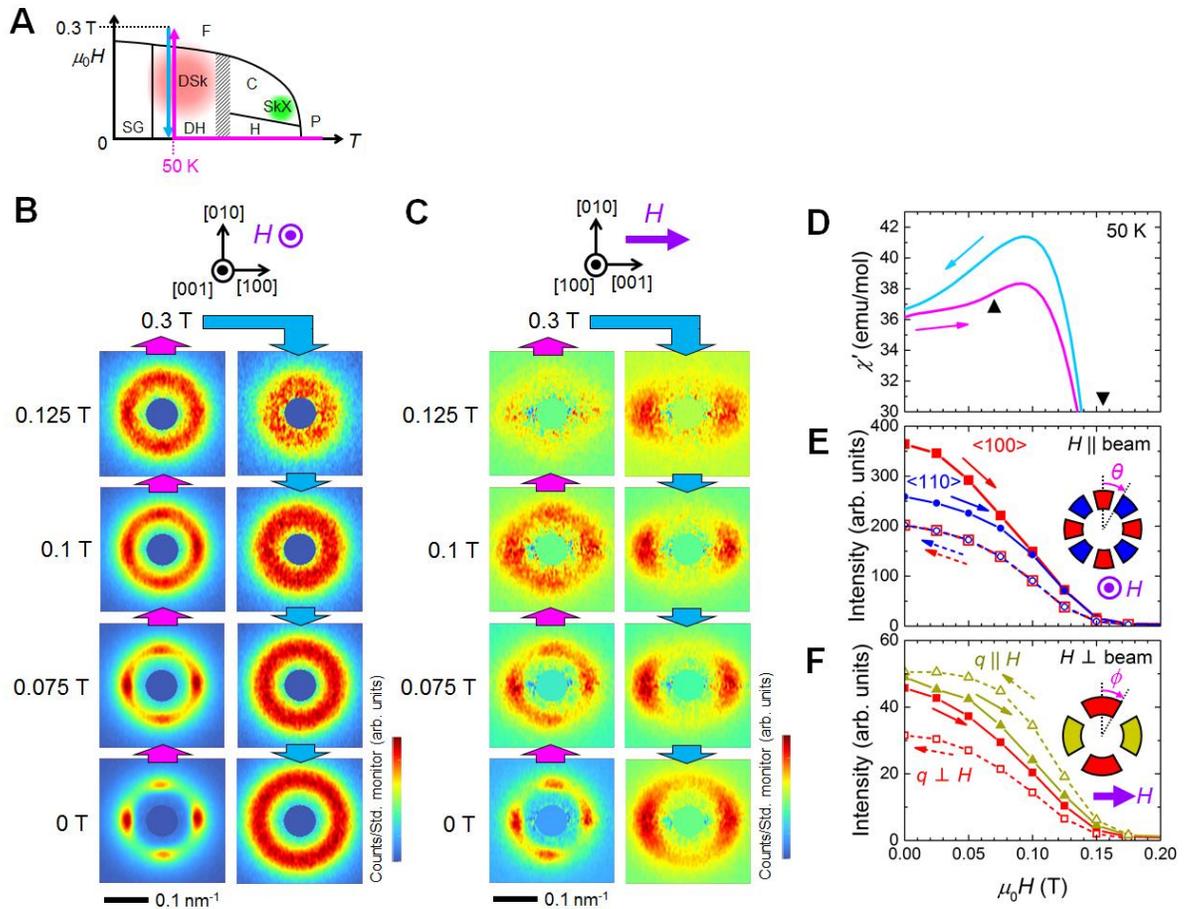

**Fig. S6. SANS and ac susceptibility measurements in the field sweeping process at 50 K in Co$_7$Zn$_7$Mn$_6$.** (**A**) Schematic of the measurement process. The field-increasing run from 0 T to 0.3 T after ZFC, and the field-decreasing run from 0.3 T to 0 T are denoted by pink and light blue arrows, respectively. (**B**) SANS images at selected fields in the $H \parallel$ beam geometry. (**C**) SANS images at selected fields in the $H \perp$ beam geometry. Note that the intensity scale of the color plots varies between each panel. (**D**) Field dependence of $\chi'$ in the field-increasing run (pink line) and decreasing run (light blue line). The phase boundaries indicated by triangles were determined similarly as in Figs. S4D,E. (**E**) Field dependence of the SANS intensities in the $H \parallel$ beam geometry integrated for directions close to <100> (red squares) and <110> (blue circles) (defined similarly as in Fig. S3F). The field-increasing and field-decreasing runs are indicated by closed and open symbols with the same colors, respectively. (**F**) Field dependence of the SANS intensities in the $H \perp$ beam geometry integrated over the regions nearly perpendicular to the field (red squares) and nearly parallel to the field (yellow triangles) (defined similarly as in Figs. S4H,I). The field-increasing and field-decreasing runs are indicated by closed and open symbols with the same colors, respectively.

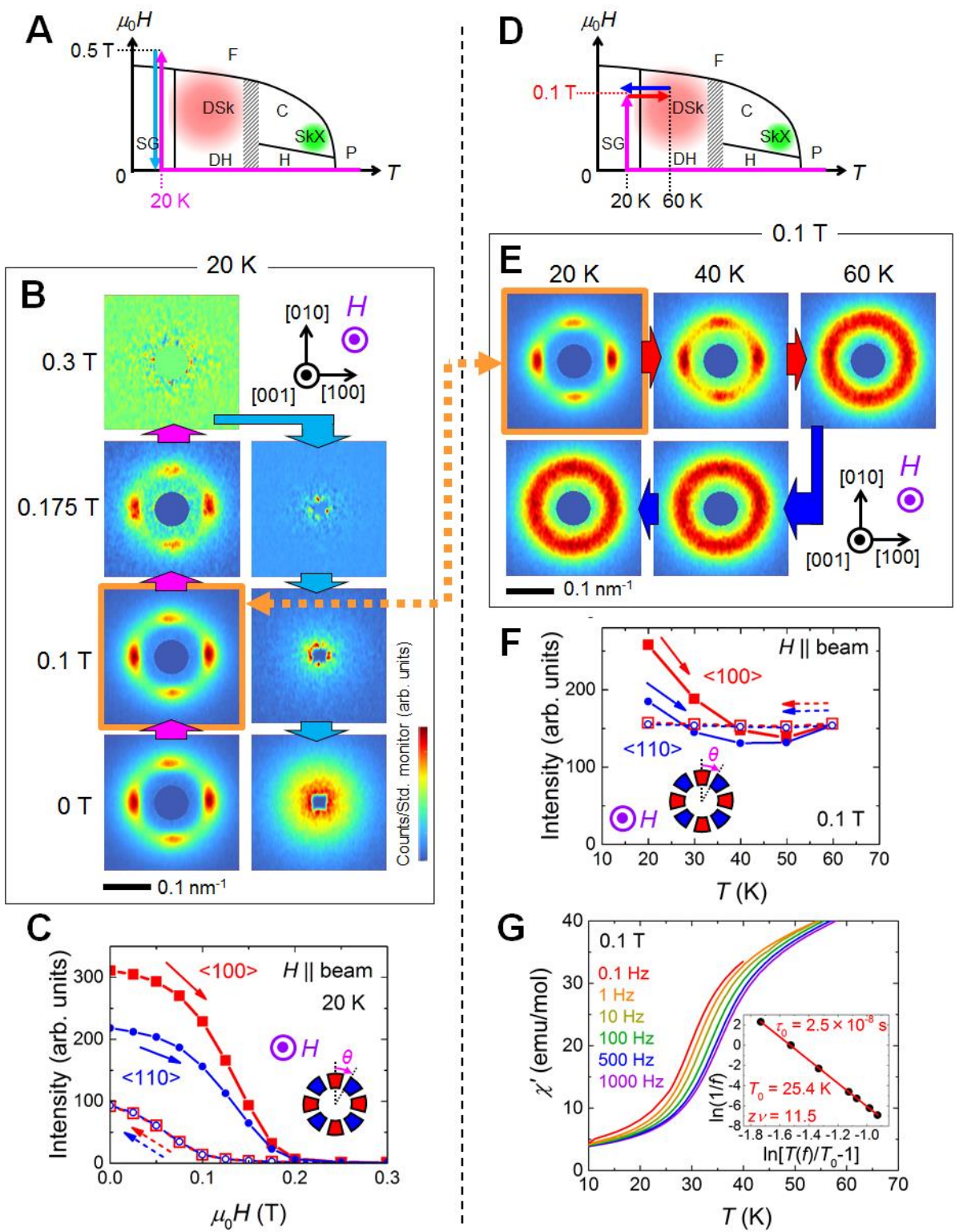

**Fig. S7. SANS and ac susceptibility measurements in the field sweeping process at 20 K and the subsequent temperature sweeping process in Co$_7$Zn$_7$Mn$_6$.** (**A**) Schematic of the measurement process for panels B and C. The field-increasing run from 0 T to 0.5 T at 20 K after ZFC, and field-decreasing run from 0.5 T to 0 T are denoted by pink and light blue arrows, respectively. (**B**) SANS images at 20 K and selected fields in the $H \parallel$ beam geometry. Note that the intensity scale of the color plots varies between each panel. (**C**) Field dependence of the SANS intensities in the $H \parallel$ beam geometry integrated for directions close to <100> (red squares) and <110> (blue circles) (defined similarly as in Fig. S3F). The field-increasing and field-decreasing runs are indicated by closed and open symbols with the same colors, respectively. (**D**) Schematic of the measurement process for panels E-G. Field-warming (FW) process from 20 K to 60 K at 0.1 T and the subsequent FC process from 60 K to 20 K are denoted by red and blue arrows, respectively. (**E**) SANS images at 0.1 T and selected temperatures in the $H \parallel$ beam geometry. Note that the intensity scale of the color plots varies between each panel. The SANS image at 20 K (warming run) is the same condition as 0.1 T (field-increasing run) in panel B as indicated by the orange frames and dotted arrow. (**F**) Temperature dependence of the SANS intensities in the $H \parallel$ beam geometry integrated for directions close to <100> (red squares) and <110> (blue circles) (defined similarly as in Fig. S3F). The FW and the FC processes are indicated by closed and open symbols with the same colors, respectively. (**G**) Temperature dependence of $\chi'$ at several ac frequencies $f$ from 0.1 Hz to 1000 Hz in the FC (0.1 T) process. The inset shows the $\ln(f^{-1})$ vs $\ln[T(f)/T_0-1]$ plot, determined similarly as in Fig. S3C. Experimental

data (black circles) are fitted to a power law (red line), $f^{-1} = \tau_0[T(f)/T_0-1]^{-zv}$, with $\tau_0 = 2.5 \times 10^{-8}$ s, $T_g = 25.4$ K and $zv = 11.5$.

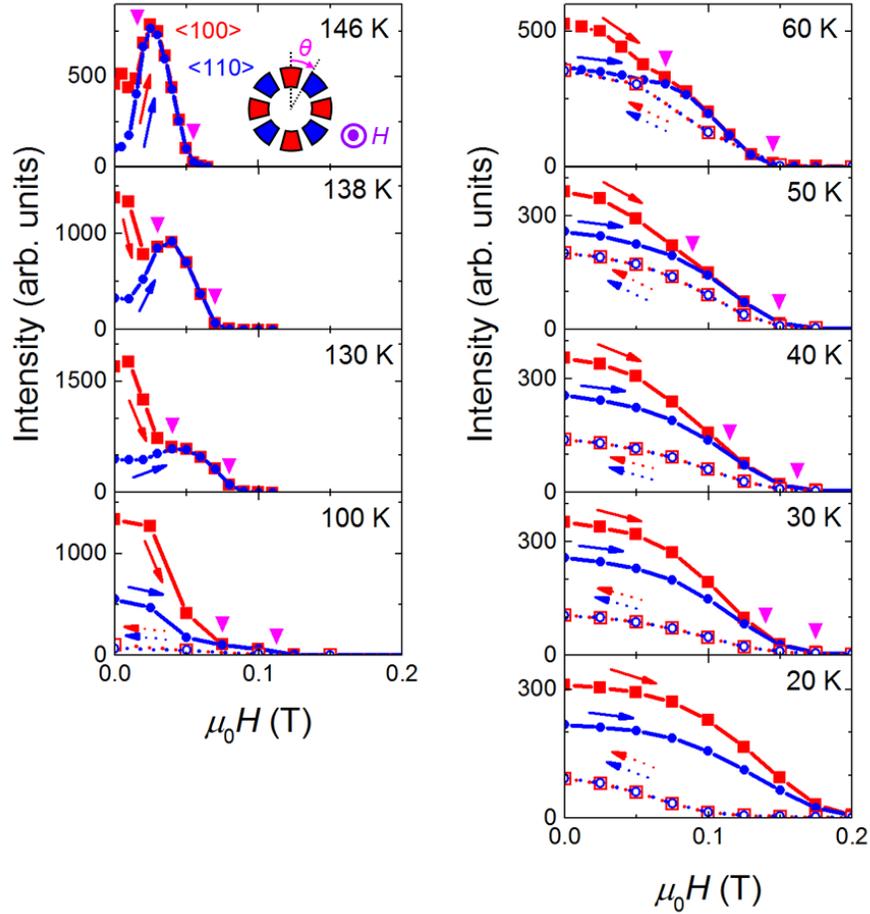

**Fig. S8. Field dependence of the SANS intensities in the $H \parallel$ beam geometry at all the measured temperatures.** These SANS intensities are integrated for directions close to <100> (red squares) and <110> (blue circles) (defined similarly as in Fig. S3F). The field-increasing and field-decreasing runs are indicated by closed and open symbols with the same colors, respectively. Pairs of pink triangles indicate the field regions where the total

intensities for directions close to <100> and <110> overlap with each other (only in the field-increasing run). These are plotted as pink circles in the *T-H* phase diagrams of Fig. 2 in the main text. The equilibrium phase of disordered skyrmions was determined as the region surrounded by the pink circles below the crossover temperature around 90 K.

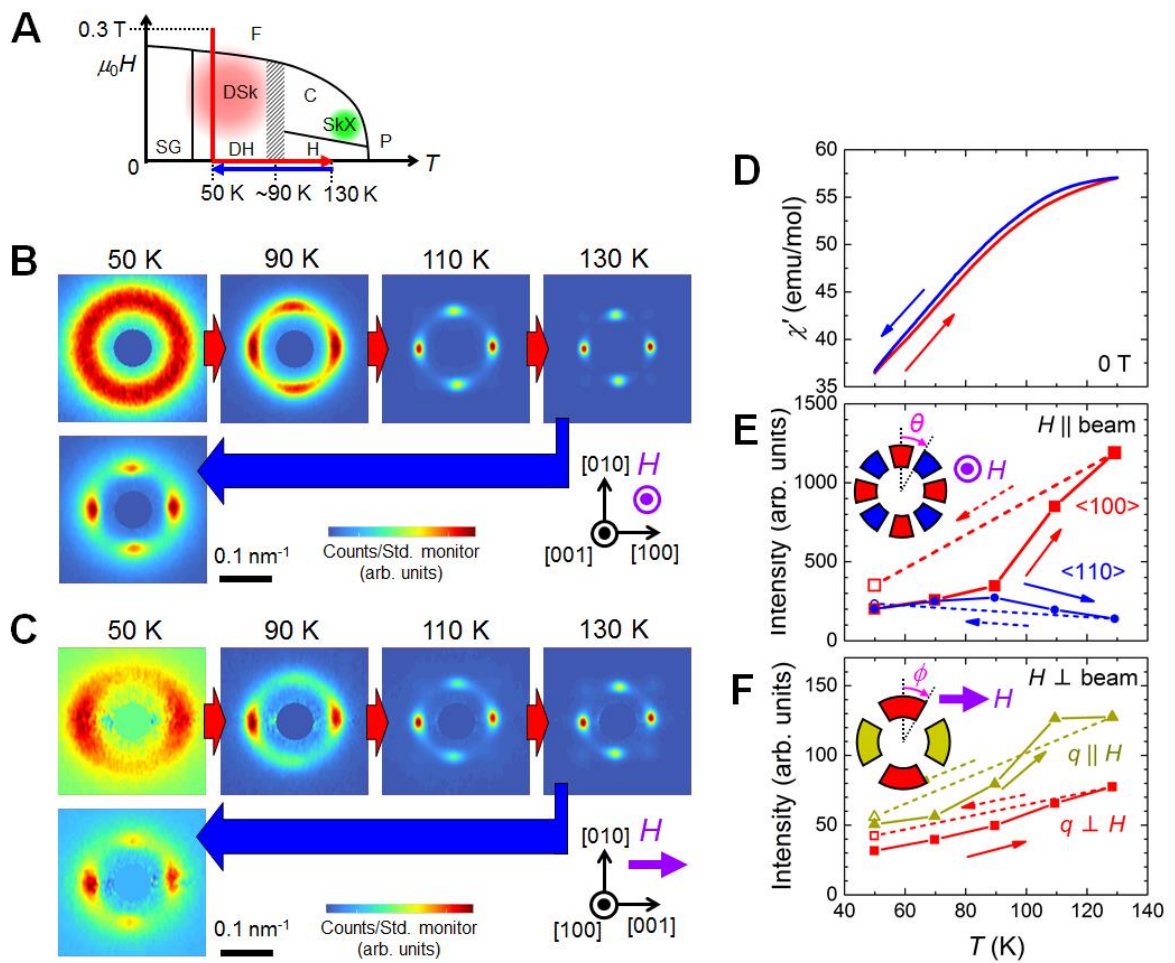

**Fig. S9. SANS and ac susceptibility measurements in the zero-field warming (ZFW) process after the field-decreasing run at 50 K, and a subsequent ZFC process, in**

$Co_7Zn_7Mn_6$. (**A**) Schematic of the measurement process. The ZFW from 50 K to 130 K after the field-decreasing run at 50 K, and the subsequent ZFC from 130 K to 50 K are indicated by red and blue arrows, respectively. (**B**) SANS images at selected temperatures in the $H \parallel$ beam geometry. (**C**) SANS images at selected temperatures in the $H \perp$ beam geometry. Note that the intensity scale of the color plots varies between each panel. (**D**) Temperature dependence of $\chi'$ in the ZFW (red line) and the ZFC (blue line) processes. (**E**) Temperature dependence of the SANS intensities in the $H \parallel$ beam geometry integrated for directions close to <100> (red squares) and <110> (blue circles) (defined similarly as in Fig. S3F). The ZFW and the ZFC processes are indicated by closed and open symbols with the same colors, respectively. (**F**) Temperature dependence of the SANS intensities in the $H \perp$ beam geometry integrated over the regions nearly perpendicular to the field (red squares) and nearly parallel to the field (yellow triangles) (defined similarly as in Fig. S4H,I). The ZFW and the ZFC processes are indicated by closed and open symbols with the same colors, respectively.

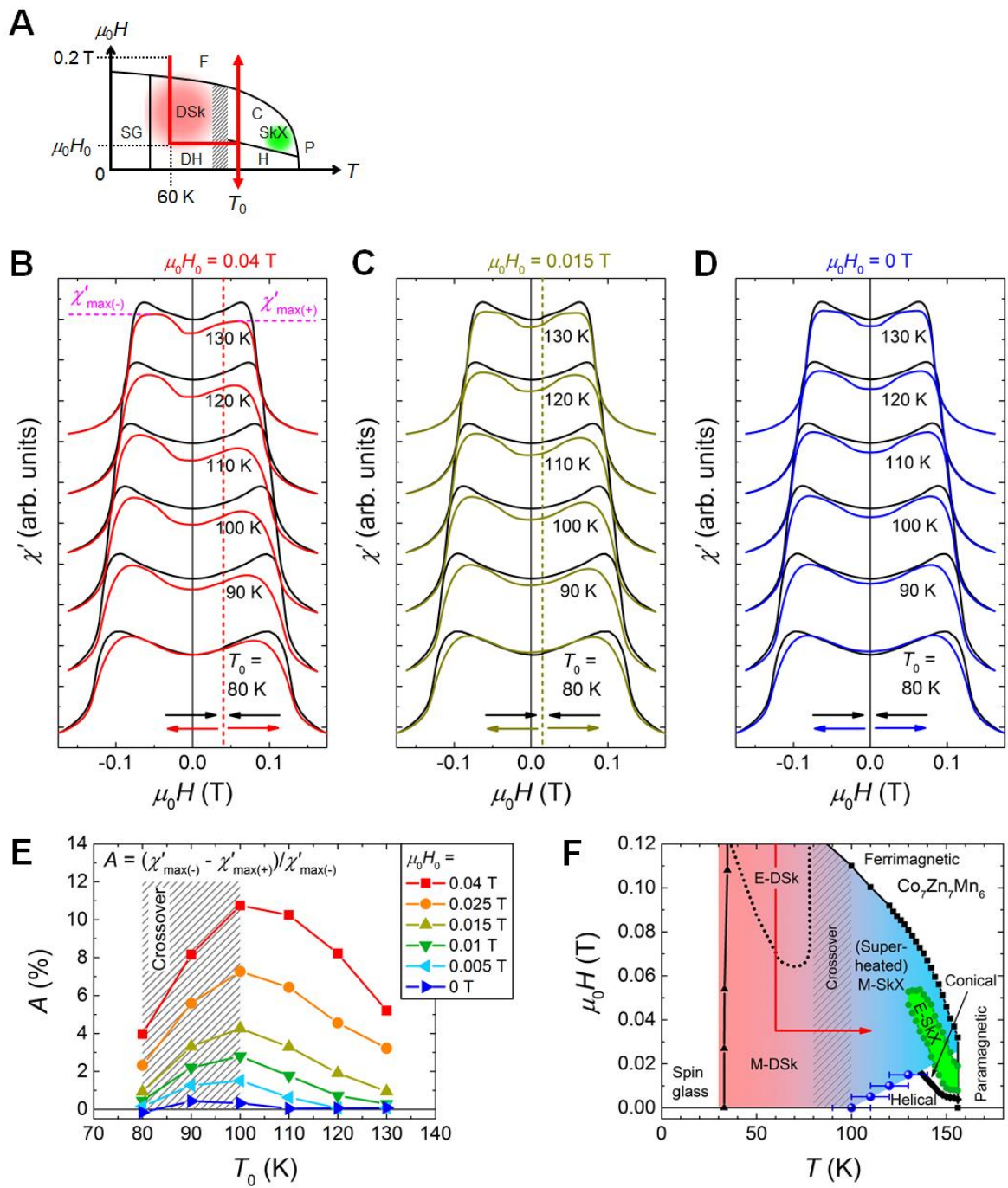

**Fig. S10. Field-swept ac susceptibility measurements after several warming processes in $Co_7Zn_7Mn_6$.** (**A**) Schematic of the measurement processes for panels B-E. (**B-D**) Field

dependence of ac susceptibility $\chi'(H)$ at several temperatures ($T_0$) done after the field decrease at 60 K and the subsequent FW with $\mu_0 H_0$ = (**B**) 0.04 T, (**C**) 0.015 T and (**D**) 0 T. Color lines show the field-sweeping (increasing $|H - H_0|$) runs. Black lines show the field-returning runs (decreasing $|H - H_0|$) from high-field ferrimagnetic regions. (**E**) Asymmetry parameter, defined as $A \equiv (\chi'_{max(-)} - \chi'_{max(+)})/\chi'_{max(-)}$, plotted against $T_0$ for various $H_0$. Here, $\chi'_{max(+)}$ and $\chi'_{max(-)}$ are local maximum values of $\chi'$ at $H > 0$ and $H < 0$, respectively, in the field-sweeping processes from $H_0$ as exemplified with dotted pink lines in panel B. The disorder-order crossover region around 90 K is shown with gray hatching. (**F**) *T-H* state diagram in several warming processes after a field decrease at 60 K (red arrow). Abbreviations of "E" and "M" indicate equilibrium and metastable, respectively. The disordered skyrmion (DSk) state, including the equilibrium phase (shown by the dotted black line) and the metastable state, is indicated by red color area. The superheated metastable SkX state is indicated by blue color area. The boundary between the metastable SkX state and the equilibrium helical/conical state (blue circles) was determined as the temperature $T_0$ where $A$ becomes less than 1% above 100 K for each $H_0$ (panel E). For the other phase boundaries, the same data points as Fig. 2A in the main text, determined by ac susceptibility in field-increasing processes after ZFC, were used.